\documentclass[lettersize,journal]{IEEEtran}
\usepackage{amsmath,amsfonts}
\usepackage{algorithm}
\usepackage{algpseudocode}
\usepackage{array}
\usepackage[caption=false,font=normalsize,labelfont=sf,textfont=sf]{subfig}
\usepackage{textcomp}
\usepackage{stfloats}
\usepackage{url}
\usepackage{verbatim}
\usepackage{graphicx}
\usepackage{cite}
\usepackage{textcase}
\usepackage{amsmath}
\usepackage{amssymb}
\usepackage[table]{xcolor}

\usepackage{booktabs}
\usepackage{multirow}
\begin{document}

\title{Memory-Coupled Unlearning for Data-Absent Identity Forgetting}

\author{Chun-I~Fan,~\IEEEmembership{Senior Member,~IEEE,}
        Lu-En~Chen,
        Ming-Feng Tsai,
        Tomohiro Morikawa
        % <-this % stops a space
\thanks{
Chun-I Fan, Lu-En Chen, and Ming-Feng Tsai are with the Department of Computer Science and Engineering, National Sun Yat-sen University, Kaohsiung 804201, Taiwan (E-mail: cifan@mail.cse.nsysu.edu.tw),
(E-mail: jack71805@gmail.com), (E-mail: parktasi@gmail.com).

Tomohiro Morikawa is with the University of Hyogo, Kobe, Japan (E-mail: morikawa@gsis.u-hyogo.ac.jp).

Corresponding author: Tomohiro Morikawa.}% <-this % stops a space
\thanks{Manuscript received ; revised }}

\markboth{Journal of \LaTeX\ Class Files,~Vol., No., }%
{Shell \MakeLowercase{\textit{et al.}}: A Sample Article Using IEEEtran.cls for IEEE Journals}


\maketitle

\begin{abstract}
Machine unlearning aims to reduce the influence of designated training
data from a trained model while preserving its utility on retained data.
Identity-level unlearning is particularly challenging when only limited
target-specific information is available when a forgetting request is
processed.
\\
In this work, we propose Memory-Coupled Unlearning (MCU), a framework
that augments a host model with a Modern Hopfield Network (MHN) serving
as a controllable external identity-prototype memory. During prototype
construction, feature representations associated with each training
identity are aggregated into identity prototypes and stored in the MHN.
A Prototype Adaptation Network (PAN) transforms the retrieved memory
representation, while a Dependency Gating Unit (DGU) controls its
contribution to downstream prediction.
\\
When an identity is requested to be forgotten, MCU first removes the
corresponding prototype from the external memory and subsequently
readapts the model using limited support information associated with
the target identities together with retained data. Importantly, in this
work, ``data-absent'' specifically refers to the absence of the complete
original forget set during request-time gradient-based readaptation;
retained data may still be used to preserve downstream-task utility.
\\
Experiments based on the identity-partitioning and one-shot support construction of 1-SHUI show that MCU achieves a favorable empirical
trade-off among model utility, membership distinguishability, and
request-time computational efficiency compared with the evaluated
baselines. These results provide empirical evidence for the potential
of controllable external memory in machine unlearning, rather than a
formal guarantee of complete information removal.
\end{abstract}


\maketitle
\IEEEdisplaynontitleabstractindextext
\IEEEpeerreviewmaketitle

\section{Introduction}
\label{ch:introduction}
\textcolor{black}{
Machine learning models, particularly deep neural networks, possess a
strong capacity for ``data memorization''~\cite{shi2024muse, yao2024machine},
which creates significant privacy risks. Attacks such as Model Inversion
and Membership Inference (MIA) have demonstrated the ability to reconstruct
recognizable information from training data or infer whether particular
samples were included in the training set~\cite{shokri2017membership,
hu2024lwywtu}. Such memorization raises important concerns under data
protection regulations such as the General Data Protection Regulation
(GDPR)~\cite{GDPR}, which establishes the ``Right to be Forgotten''
(RTBF)~\cite{RTBF, correia2021right, kelly2021achieve, kelly2019gdpr}.
\\
In response to increasing concerns regarding the retention and use of
personal data, governments worldwide have introduced data-protection
regulations~\cite{teixeira2023can, novikovas2022employee}. In particular,
Article 17 of the GDPR provides individuals with the right to request the
deletion of their personal data~\cite{correia2021right, kelly2021achieve,
kelly2019gdpr}. Together with related regulations such as the CCPA in
California and PIPL in China~\cite{goldfield2020right, zhen2022existence},
these requirements create practical challenges for trained machine learning
models because information derived from training samples may remain
distributed across model parameters~\cite{xu2024machine, thudi2022necessity,
eisenhofer2025verifiable}.
\\
The most direct reference solution for machine unlearning is model
retraining, in which the model is trained again after excluding the data
associated with the requested identities. Although retraining provides an
important empirical reference, it can incur substantial computational cost
because model optimization must be repeated using the retained training
data. Moreover, in practical unlearning scenarios, the complete original
data associated with the identities requested to be forgotten may no
longer be available when a forgetting request is processed. These
considerations motivate efficient unlearning mechanisms that reduce the
influence of requested identities without requiring direct reuse of their
complete original training samples during model readaptation.
\\
In this paper, we explicitly distinguish data availability across different
stages of the unlearning lifecycle. Let $\mathcal{D}_{f}$ denote the
complete original forget set associated with the identities requested to
be forgotten, $\mathcal{D}_{r}$ denote the retain set, and
$\mathcal{D}_{\text{support}}$ denote the limited support information
provided for the requested identities. During initial model training and
prototype construction, the original training data are available. After
the host model has been trained, its feature extractor is used to construct
identity-associated prototypes that are stored in an external memory.
\\
When a subsequent forgetting request is processed, the complete original
forget set $\mathcal{D}_{f}$ is not directly revisited for gradient-based
model readaptation. Instead, MCU uses the limited support information in
$\mathcal{D}_{\text{support}}$ together with the retain set
$\mathcal{D}_{r}$ to guide model readaptation and preserve downstream-task
utility. Accordingly, MCU should not be interpreted as a fully data-free
or retain-data-free unlearning pipeline. Throughout this paper, the term
\textit{data-absent} refers specifically to the absence of the complete
original forget data during request-time gradient-based readaptation,
rather than to the absence of all training-derived information or retained
data. This distinction is made explicitly to avoid ambiguity regarding the
data-access assumptions of the proposed framework.
\\
This setting motivates an \textbf{unlearning-by-design} perspective,
in which explicit support for subsequent forgetting is incorporated
through a controllable external-memory mechanism before request-time
unlearning is performed. Rather than relying solely on post-hoc modification of model
parameters, MCU augments a host model with an external identity-prototype
memory together with a PAN and a DGU. The external memory provides explicitly manageable
identity-associated entries, while PAN and DGU mediate the interaction
between the retrieved memory representation and the host-model
representation.
\\
Importantly, the term \textit{unlearning-by-design} in this work does not
imply that the host model must be trained jointly with the memory-coupling
components from scratch. Instead, it refers to the requirement that the
external prototype memory and its interaction mechanism be prepared as
part of the MCU pipeline before request-time forgetting is performed.
Consequently, MCU is not a purely parameter-space drop-in unlearning
method that can be applied to an arbitrary deployed model without
additional preparation. This requirement represents a trade-off between
direct post-hoc applicability and explicit controllability of
identity-associated external memory.
\\
Building on this perspective, we propose the MCU framework. MCU employs a MHN as a differentiable external memory for storing
identity-associated prototypes. When an identity is requested to be
forgotten, MCU first removes the corresponding prototype from the external
memory and then performs lightweight gradient-based readaptation of the
host and memory-interaction components using
$\mathcal{D}_{\text{support}}$ together with $\mathcal{D}_{r}$.
The complete original forget set $\mathcal{D}_{f}$ is not directly used
during this readaptation stage.
\\
In the formulation considered in this work, \textit{identity} and
\textit{downstream-task target} play different roles. Identity determines
the unit to which a forgetting request applies, whereas the downstream-task
label defines the prediction objective of the host model. Therefore,
identity-level unlearning in this work does not imply that the host model
itself performs identity classification. For example, the downstream
prediction task may involve facial attributes or age groups while the
forgetting request is specified at the identity level.
\\
Because each stored prototype is constructed from feature representations
derived from samples associated with a particular identity, we
conservatively regard the prototype as a
\textbf{privacy-sensitive derived representation}. Therefore, MCU does
not assume that storing prototypes is equivalent to eliminating all
information derived from the original training data, nor do we claim that
the prototypes are inherently free from identity-related information.
Instead, the purpose of the external memory is to provide an explicit and
controllable representation whose corresponding entry can be directly
removed when a forgetting request is received.
}
% The most direct solution, model retraining, is often computationally prohibitive~\cite{chien2024langevin, bourtoule2021machine, yao2024machine}. For large models, retraining can take days or weeks, consuming substantial resources~\cite{xu2024machine, sepahvand2024data}. To address this dilemma, Machine Unlearning has emerged, aiming to selectively "forget" the influence of specific data without retraining the entire model~\cite{thudi2022necessity, xu2025machine}. This field represents an important shift in machine learning research, from purely pursuing performance to balancing utility with privacy and social responsibility~\cite{xu2024machine, geng2025comprehensive}.

% However, a critical challenge paralyzes most existing unlearning algorithms: the \textbf{data-absent scenario}~\cite{sepahvand2024data, de2025unlearning}. Due to legal or practical constraints, the original training data is often unavailable when an unlearning request is made. This paper tackles this data-absent challenge not through algorithmic refinement, but through architectural innovation. We propose the \textbf{Memory-Coupled Unlearning (MCU)} framework, a novel approach that integrates a differentiable external memory to achieve efficient and precise unlearning without relying on original data.
\textcolor{black}{
The primary contributions of this work are summarized as follows:
}
\begin{itemize}

 \item We introduce an MHN-based differentiable external memory
    for identity-level machine unlearning. The external memory provides
    explicitly manageable identity-associated prototype entries that can
    be directly removed when a forgetting request is processed. During
    request-time readaptation, the complete original forget set is not
    directly revisited, while limited support information and retained
    data are used for model readaptation.
    \item We propose MCU, an unlearning-by-design framework that couples
    a host model with a controllable external identity-prototype memory.
    The prototype memory and its interaction mechanism are prepared
    before request-time forgetting, enabling explicit prototype removal
    followed by lightweight model readaptation.
    \item We design a PAN, a DGU, and a dependency-inducing objective
    to encourage interaction with the external-memory pathway. Together
    with the retain-utility objective, these components support
    readaptation after prototype removal while preserving downstream-task
    utility on retained data.
    \item We empirically evaluate MCU using the identity-partitioning and
    one-shot support construction of the 1-SHUI benchmark. Across the
    evaluated settings, MCU provides a favorable empirical trade-off
    among downstream-task utility, membership distinguishability, and
    computational efficiency, including up to a 54.8-fold request-time
    speedup relative to retraining after prototype preparation.
    \item We introduce the Utility-Privacy Balance Index (UPBI) as a
    complementary descriptive metric for summarizing the observed
    utility--membership trade-off. UPBI is intended to complement,
    rather than replace, the individual utility and attack-specific
    privacy measurements.

\end{itemize}

\textcolor{black}{
Overall, MCU provides an external-memory-based perspective on
identity-level machine unlearning by combining explicit prototype
management with lightweight model readaptation.
}
\section{Background and Related Work}
\label{ch:relatedworkandpreliminaries}

This section first establishes the theoretical foundation of Modern Hopfield Networks (MHN), which is central to our proposed architecture. It then provides a systematic review of existing machine unlearning methodologies, analyzing their core mechanisms and data dependencies to position our work within the broader landscape.

\subsection{Modern Hopfield Networks as External Memory}

\subsubsection{Theoretical Foundation: Energy and Update}
The core mechanism of a Modern Hopfield Network (MHN) is defined by its energy function $E(\boldsymbol{\xi})$ for a state pattern $\boldsymbol{\xi}$. Deep learning, Ramsauer et al.~\cite{ramsauer2020hopfield} formulated this energy for continuous patterns stored in a matrix $X$ using the Log-Sum-Exp (LSE) function:
\begin{equation}
    E(\boldsymbol{\xi}) = -\text{lse}(\beta, X^T \boldsymbol{\xi}) + \frac{1}{2}\boldsymbol{\xi}^T\boldsymbol{\xi} + C
\end{equation}
where $X \in \mathbb{R}^{d \times N}$ represents the memory of $N$ stored patterns, $\beta$ is a temperature parameter, and lse is the Log-Sum-Exp function. The state update rule that minimizes this energy is given by:
\begin{equation}
    \boldsymbol{\xi}^{new} = X \text{softmax}(\beta X^T \boldsymbol{\xi})
\end{equation}
This update rule is mathematically equivalent to the attention mechanism in Transformers, where the state vector serves as the query, and the stored patterns serve as both keys and values.

\subsubsection{MHN as a Differentiable Unlearning Interface}

A useful property of MHNs in the context of MCU is that their stored
patterns are explicitly represented and can be integrated into
differentiable neural-network pipelines. This enables the external
prototype memory to serve as a controllable interface for subsequent
unlearning operations.

When a stored prototype is explicitly removed from the MHN memory,
that prototype no longer participates in subsequent memory retrieval.
This operation removes the corresponding entry from the external
prototype store, but does not by itself imply that all information
associated with the corresponding identity has been removed from the
host-model parameters.

\subsection{A Taxonomy of Machine Unlearning Methods}
%\textcolor{black}{
Machine unlearning research has increasingly emphasized not only
computational efficiency, but also careful evaluation of the residual
privacy risks after unlearning. Recent critiques by Hayes
et al.~\cite{hayes2025inexact} and Naderloui
et al.~\cite{naderloui2025rectifying} have highlighted that
approximate unlearning methods may provide an incomplete picture of
privacy when evaluated only under limited or non-adaptive attack
settings. These observations motivate the development of unlearning
mechanisms that provide more explicit and controllable forgetting
operations, together with empirical evaluation under clearly defined
threat models. Such controllability should not be interpreted as a
formal guarantee of complete information removal. Based on their
technical mechanisms, we categorize existing methods as follows:
\subsubsection{Data Partitioning and Isolated Retraining}
Data-partitioning approaches organize the training data and model
training process so that only a subset of the model needs to be
retrained when an unlearning request is received. The SISA framework is
a representative example that partitions the training data into
multiple shards and slices to reduce the cost of subsequent
retraining~\cite{bourtoule2021machine}. Although such approaches provide
a clear retraining-based mechanism for removing the influence of
specified data, they introduce additional storage and model-management
overhead and require the corresponding partitioning structure to be
maintained throughout the model lifecycle.
\subsubsection{Approximate Unlearning via Parameter-Space Modification}
These methods modify model parameters directly to reduce the influence
of the requested data while attempting to preserve useful
representations for the remaining data
~\cite{shi2025deepclean, foster2024loss, yamashita2023one}.
Fisher Information Matrix (FIM)-based approaches are representative
examples. DeepClean, for instance, utilizes FIM diagonal elements to
identify and selectively update parameters that are sensitive to the
target data
~\cite{shi2025deepclean, iacovelli2022gwfast, neilsen2018optimal}.
Other methods in this category include Selective Synaptic Dampening
(SSD)~\cite{foster2024fast}. The request-time data requirements of these
parameter-space approaches vary across methods; depending on the
specific procedure, they may rely on target forgetting data, retained
data, or other auxiliary information. Their forgetting behavior is
therefore typically assessed empirically under the corresponding
data-access and threat-model assumptions
~\cite{shi2025deepclean, fan2023salun}.
\\
\subsubsection{Decision Boundary Adjustment Methods}
These methods perform unlearning by modifying the decision boundary of
the trained model rather than reconstructing the model from scratch
~\cite{chen2023boundary, lee2023undo, shrivastava2024refining}.
Boundary Unlearning, for example, applies boundary contraction and
expansion operations to approximate the behavior of a model retrained
without the requested data
~\cite{chen2023boundary, chen2024machine}. Such approaches can reduce
the computational cost of unlearning, although their request-time data
requirements depend on the specific boundary-adjustment procedure.
Some methods require access to retained or other auxiliary data during
the adjustment process.
\\
\subsubsection{Knowledge Transfer and Generative Model Methods}
Another line of work employs knowledge distillation, adversarial
training, or generative mechanisms to transfer retained knowledge while
reducing the influence of the requested data
~\cite{kurmanji2023towards, fan2023salun, lin2023erm}. The Bad Teacher
method, for instance, adopts a teacher--student framework in which the
student receives different supervisory signals for retained and
forgotten data~\cite{chundawat2023can}. These approaches provide
flexible mechanisms for approximate unlearning, but their request-time
data-access assumptions vary considerably and may involve retained
data, forget data, generated samples, or combinations of these sources.
\subsubsection{Data Availability in Machine Unlearning}
The terms \textit{data-free} and \textit{data-absent} have been used
with different assumptions regarding data availability across the
machine-unlearning lifecycle. To avoid ambiguity, we distinguish the
data required during initial model training and prototype construction
from those available during request-time model readaptation.
\\
In the setting considered in this work, the original training data are
available during initial model training and prototype construction.
These data are used to train the host model and to construct the
identity-associated prototypes stored in the external memory. When a
subsequent forgetting request is processed, the complete original forget
set $\mathcal{D}_{f}$ is not directly revisited for gradient-based
readaptation. Instead, the limited support information associated with
the requested identities, denoted by
$\mathcal{D}_{\text{support}}$, together with the retain set
$\mathcal{D}_{r}$, is available during model readaptation.
\\
Accordingly, MCU should not be interpreted as a fully data-free or
retain-data-free unlearning method. Throughout this work, the term
\textit{data-absent} specifically refers to the absence of the complete
original forget data during request-time gradient-based readaptation,
rather than to the absence of all training-derived information or
retained data. This distinction provides a more precise characterization
of the data-access assumptions considered by MCU.
\subsubsection{Positioning Our Work}
Existing machine unlearning methods commonly perform post-hoc
modification of a trained model through parameter updates, gradient
manipulation, knowledge transfer, or related mechanisms. In contrast,
MCU introduces an explicitly manageable external prototype memory as an
additional mechanism for identity-level unlearning.
\\
Specifically, MCU augments a host model with a MHN, a PAN, and a DGU. Identity-associated prototypes are constructed and stored
in the external memory before request-time forgetting is performed.
When a forgetting request is received, the prototype corresponding to
the requested identity can be explicitly removed, after which the host
model and memory-interaction components are readapted to the modified
memory state using the support and retain information available at
request time.
\\
This design follows an \textbf{unlearning-by-design} perspective in the
sense that an explicit mechanism for subsequent forgetting is prepared
before request-time unlearning is performed. Importantly, this does not
imply that the host model must be jointly trained with the memory-coupling
components from scratch. Rather, the external prototype memory and its
interaction mechanism are incorporated into the MCU pipeline before a
forgetting request is processed.
\\
Consequently, MCU is not a purely parameter-space drop-in unlearning
method that can be applied to an arbitrary deployed model without
additional preparation. Instead, it represents a trade-off between
direct post-hoc applicability and the explicit controllability of
identity-associated external memory.
%}
\section{The Memory-Coupled Unlearning (MCU) Framework}
\label{ch:methodology}
%\textcolor{black}{
Building upon the challenges of the \textbf{unlearning-time data-absent}
setting and ``one-shot unlearning,'' this section introduces the proposed MCU framework. MCU follows an
\textbf{unlearning-by-design} perspective in which an explicit
external-memory mechanism for subsequent forgetting is prepared before
request-time unlearning is performed. Specifically, MCU augments a host
neural network with an external MHN that stores identity-associated
prototypes, together with a PAN and a DGU that mediate the interaction
between the retrieved memory representation and the host-model
representation.
This memory-coupled design provides an explicitly manageable pathway
through which identity-associated prototype entries can be removed when
a forgetting request is received. During subsequent model readaptation,
the modified external memory interacts with the host representation
through the PAN and DGU. The purpose of this design is to increase the
controllability of identity-associated external memory and the influence
of explicit prototype removal on the unlearning process. Importantly,
this design does not assume or guarantee that all information associated
with an identity is stored exclusively in the external memory.
%}
\subsection{Problem Formulation}
\label{subsec:problem-formulation}
%\textcolor{black}{
Let the original training set be
\begin{equation}
    \mathcal{D}_{tr}
    =
    \{(x_j,i_j,t_j)\}_{j=1}^{N},
\end{equation}
where $x_j$ denotes an input sample, $i_j$ denotes its identity
identifier, and $t_j$ denotes its downstream-task target.
Importantly, the identity identifier and the downstream-task target play
different roles in our formulation. The identity identifier $i_j$
determines the unit to which a forgetting request applies, whereas
$t_j$ defines the prediction objective of the host model. Therefore,
identity-level unlearning in this work does not imply that the host
model itself performs identity classification. For example, an identity
may specify the person to be forgotten while the downstream task
corresponds to facial-attribute prediction or age-group classification.
\\
Let $\mathcal{I}_{tr}$ denote the set of identities represented in the
training data, and let
$\mathcal{I}_{f}\subseteq\mathcal{I}_{tr}$ denote the set of identities
requested to be forgotten. The corresponding forget set is defined as
\begin{equation}
    \mathcal{D}_{f}
    =
    \{(x,i,t)\in\mathcal{D}_{tr}
    \mid i\in\mathcal{I}_{f}\},
\end{equation}
while the retain set is
\begin{equation}
    \mathcal{D}_{r}
    =
    \mathcal{D}_{tr}\setminus\mathcal{D}_{f}.
\end{equation}
Following the one-shot support construction, one held-out support sample
is provided for each requested identity. We denote the resulting support
set as
\begin{equation}
    \mathcal{D}_{\mathrm{support}}
    =
    \{(x_i^{s},i,t_i^{s})
    \mid i\in\mathcal{I}_{f}\},
\end{equation}
where $x_i^{s}$ is the support sample associated with identity $i$,
and $t_i^{s}$ denotes its downstream-task target. The identity
identifier $i$ is used to identify the corresponding prototype entry
in the external memory when the forgetting request is processed.
The data used at different stages of MCU are explicitly distinguished.
During initial model training and prototype construction, the original
training data are available. At request time, the complete original
forget set $\mathcal{D}_{f}$ is not directly revisited for
gradient-based readaptation. Instead, MCU uses
$\mathcal{D}_{\mathrm{support}}$ together with the retain set
$\mathcal{D}_{r}$. Thus, the data-absent assumption considered in this
work specifically concerns the absence of the complete original forget
data during request-time gradient-based readaptation.
\\
The objective of identity-level unlearning is to reduce the influence
associated with the identities in $\mathcal{I}_{f}$ while preserving
the downstream-task utility associated with retained data. We use a
model retrained without the forgotten identities as an empirical
reference for evaluating the behavior of the unlearned model. However,
we do not assume or claim formal statistical indistinguishability
between the MCU-unlearned model and the retrained model.
%}
\subsection{Framework Overview}
\begin{figure}[h!]
    \centering
    \includegraphics[width=1\linewidth]{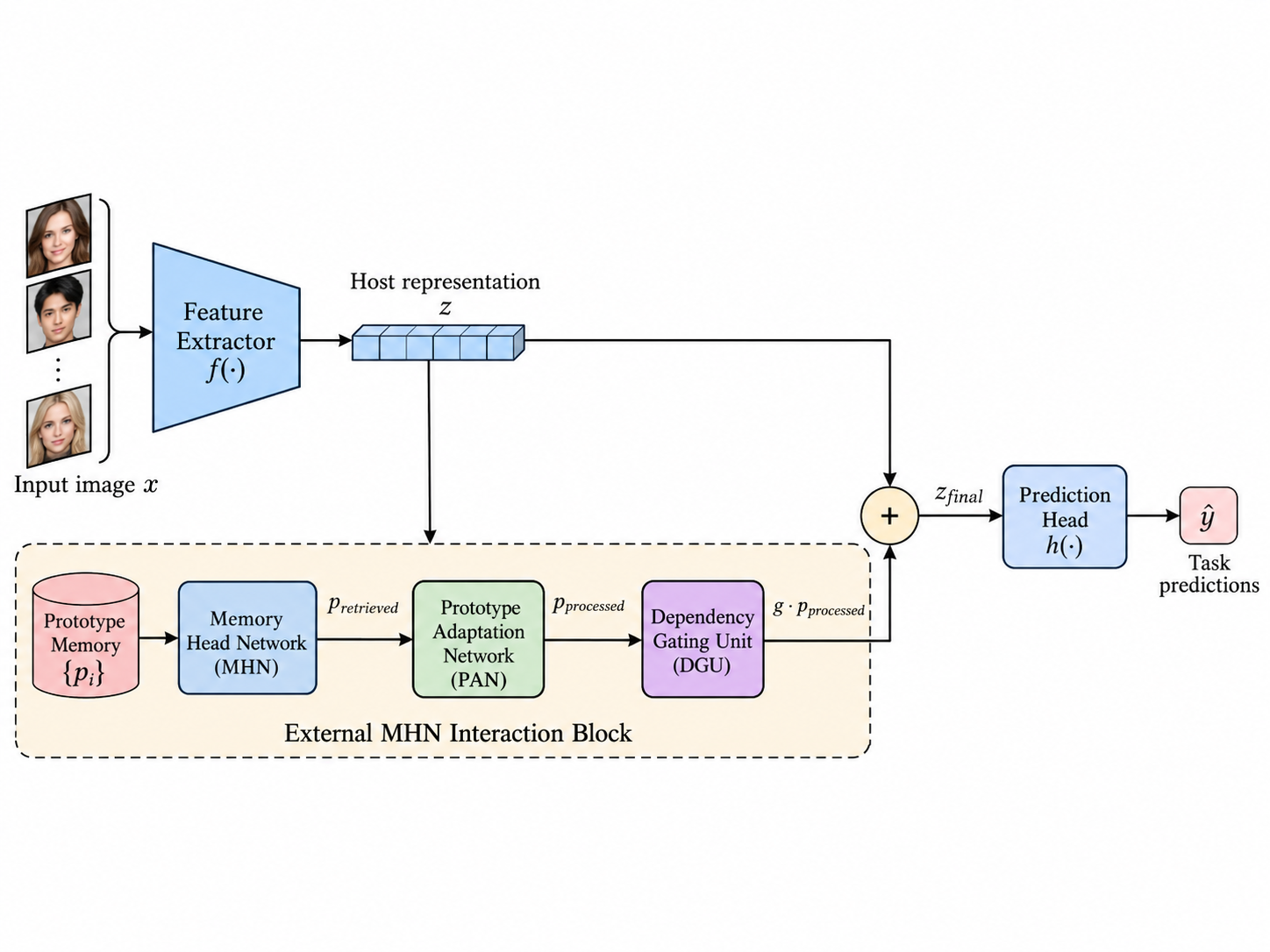}
    \caption{
        \textbf{The MCU Framework Architecture.}
 The framework consists of a
standard host model, comprising a feature extractor and a prediction
head, together with an External MHN Interaction Block containing the
MHN, PAN, and DGU. The interaction block establishes an explicit
memory-interaction pathway through which the retrieved prototype
representation is integrated with the host representation before
downstream prediction.
    }
    \label{fig:mcu-framework}
\end{figure}

%\textcolor{black}{
The MCU framework is built upon a host neural network, which typically
consists of a \textbf{feature extractor} ($f_{\text{feat}}$) and a
\textbf{classification head} ($f_{\text{head}}$). MCU augments the host
model with an external MHN, which serves as a differentiable and
controllable memory for storing identity-associated prototypes prepared
before request-time unlearning. The interaction between the host model
and the external memory is mediated by two purpose-built, lightweight
modules: the PAN and the DGU. PAN transforms the prototype representation
retrieved from the MHN into a representation compatible with the host
feature space, whereas DGU dynamically controls the contribution of the
external-memory representation to the final representation used for
downstream prediction.
Together, these components establish an explicit memory-interaction
pathway between the external prototype memory and the host representation.
This pathway enables changes in the external memory, such as explicit
prototype removal, to affect subsequent model readaptation and downstream
prediction. The purpose of this design is to improve the controllability
of identity-associated external memory rather than to imply that the host
model is formally dependent on the external memory or that all
identity-related information is stored exclusively in it. The complete
architecture is presented in Figure~\ref{fig:mcu-framework}.
%}

\subsection{Constructing the External Memory: The Identity Prototype Store}
\label{subsec:constructing-memory}

%\textcolor{black}{
Within the MCU framework, the MHN serves as a differentiable external
associative memory that stores an \textbf{identity prototype} for each
identity represented in the data used for prototype construction. Each
prototype is a compact feature representation derived from samples
associated with a particular identity. Prototype construction is
performed after the host feature extractor has been trained and before
gradient-based MCU readaptation is performed.
Importantly, the original training data are available during this
prototype-construction stage. During request-time readaptation, however,
the complete original forget set $\mathcal{D}_{f}$ is not directly
revisited. Instead, MCU uses the support set
$\mathcal{D}_{\text{support}}$ together with the retain set
$\mathcal{D}_{r}$, as defined in the preceding problem formulation.
The prototype construction procedure consists of the following steps:
%}

\begin{enumerate}

    \item \textbf{Feature Extraction:}
    The trained feature extractor $f_{\text{feat}}$, operated in
    evaluation mode during prototype construction, maps each input
    sample $x_j$ to a feature representation
    \begin{equation}
        z_j = f_{\text{feat}}(x_j).
    \end{equation}

    \item \textbf{Prototype Aggregation:}
    For each identity $i$, the feature representations associated with
    that identity are aggregated by their mean:
    \begin{equation}
        p_i =
        \frac{1}{N_i}
        \sum_{j=1}^{N_i}
        f_{\text{feat}}(x_j),
    \end{equation}
    where $N_i$ denotes the number of prototype-construction samples
    associated with identity $i$.

    \item \textbf{Memory Storage:}
    The resulting identity prototypes are normalized and stored as the
    patterns of the MHN, forming an external content-addressable
    prototype memory.
\end{enumerate}

%\textcolor{black}{
Because each prototype is derived from feature representations
associated with a particular identity, we conservatively regard the
stored prototype as a \textbf{privacy-sensitive derived representation}.
MCU therefore does not assume that the prototype memory is inherently
free from identity-related information, nor does prototype storage imply
that all information derived from the original training data has been
removed.
Once constructed, the remaining prototypes are kept fixed during
gradient-based model readaptation and are not updated through
backpropagation. Memory contents are modified only through explicit
memory-management operations. When an identity is requested to be
forgotten, its corresponding prototype is removed from the external
memory. This operation removes the target identity's explicit prototype
entry, but does not by itself imply that all information associated
with that identity has been eliminated from the host-model parameters.
MCU therefore combines explicit prototype removal with subsequent
gradient-based model readaptation.
%}

\subsection{Core Mechanism: Architectural Coupling via PAN and DGU}
\label{subsec:core-mechanism}

%\textcolor{black}{
The core of MCU is the ``External MHN Interaction Block,'' which
mediates the interaction between the host representation and the
external prototype memory. The block consists of the Modern Hopfield
Network, a \textbf{PAN}, and a
\textbf{DGU}.
Given an input representation, the MHN first performs content-based
association over the stored prototypes. The retrieved memory
representation is then transformed by PAN, while DGU determines the
magnitude of its residual contribution to the host representation.
%}

\subsubsection{MHN Association and Prototype Retrieval}
%\textcolor{black}{
When the external MHN is initialized, each identity prototype
$\mathbf{p}_i$ is $\ell_2$-normalized and stored as
$\bar{\mathbf{p}}_i$. Given a host representation $\mathbf{z}$ as a
query, MCU similarly applies $\ell_2$ normalization before association:
\begin{equation}
\bar{\mathbf{p}}_i
=
\frac{\mathbf{p}_i}{\|\mathbf{p}_i\|_2},
\qquad
\bar{\mathbf{z}}
=
\frac{\mathbf{z}}{\|\mathbf{z}\|_2}.
\end{equation}
\\
The association weight for the $i$-th stored prototype is computed as
\begin{equation}
a_i
=
\frac{
\exp\left(
\beta\,\bar{\mathbf{z}}^{\top}\bar{\mathbf{p}}_i
\right)
}{
\sum_j
\exp\left(
\beta\,\bar{\mathbf{z}}^{\top}\bar{\mathbf{p}}_j
\right)
},
\end{equation}
where $\beta$ denotes the inverse-temperature parameter controlling
the sharpness of the association distribution.
\\
The retrieved memory representation is then
\begin{equation}
\mathbf{p}_{\mathrm{retrieved}}
=
\sum_i a_i \bar{\mathbf{p}}_i.
\end{equation}
\\
Thus, the MHN retrieves a weighted combination of the available
prototype patterns rather than selecting a single identity-specific
prototype. After a target prototype is removed, subsequent association
is performed only over the remaining prototypes.
%}
\subsubsection{Prototype Adaptation Network (PAN)}

%\textcolor{black}{
The PAN serves as a learnable representation-adaptation module. It
transforms the memory representation retrieved by the MHN into a
representation compatible with the host feature space:
\begin{equation}
    p_{\text{processed}}
    =
    \operatorname{ReLU}
    \left(
        W_{\text{pan}}p_{\text{retrieved}}
        +
        b_{\text{pan}}
    \right),
\end{equation}
where $W_{\text{pan}}$ and $b_{\text{pan}}$ denote the trainable
parameters of PAN.
\\
PAN does not determine which identity should be forgotten. Its role is
to transform the content retrieved from the modified external memory
before that information is combined with the host representation.
%}

\subsubsection{Dependency Gating Unit (DGU)}
%\textcolor{black}{
The DGU determines the magnitude of the external-memory contribution
based on the host representation $z$. Specifically, it computes a
scalar gating value $g\in[0,1]$ as
\begin{equation}
    g
    =
    \sigma
    \left(
        W_{\text{dgu}}z+b_{\text{dgu}}
    \right),
\end{equation}
where $\sigma(\cdot)$ denotes the sigmoid function, and
$W_{\text{dgu}}$ and $b_{\text{dgu}}$ denote the trainable DGU
parameters.
The gating value $g$ controls the magnitude of the additional
memory-derived contribution that is subsequently incorporated into the
host representation through the normal memory-coupled path. A larger
value of $g$ increases the contribution of the external-memory pathway,
whereas a smaller value reduces this contribution. The DGU should
therefore be interpreted as a residual memory-gating mechanism rather
than as a selector that determines which identity is forgotten.
%}

\subsubsection{Dual-Path Forward Propagation}
\label{subsubsec:dual-path}
%\textcolor{black}{
During request-time readaptation, MCU employs a normal
memory-coupled path and a memory-deprived path. The two paths are used
to jointly preserve downstream-task behavior on the memory-coupled
representation and regulate the predictive capability of the
host-only representation for the support samples associated with the
requested identities.
%}
\\

\begin{enumerate}

    \item \textbf{Normal Path (Memory-Coupled):}

    Given an input $x$, the host feature extractor first produces
    \begin{equation}
        z=f_{\text{feat}}(x).
    \end{equation}

    The representation $z$ is used as a query to the MHN, producing
    $p_{\text{retrieved}}$ through the soft association described
    above. PAN then produces
    $p_{\text{processed}}$, and DGU computes $g$ from $z$.

    The final memory-coupled representation is
    \begin{equation}
        z_{\text{final}}
        =
        z
        +
        g\,p_{\text{processed}},
    \end{equation}
    and the corresponding downstream prediction is
    \begin{equation}
        y_{\text{real}}
        =
        f_{\text{head}}(z_{\text{final}}).
    \end{equation}

    \item \textbf{Ablated Path (Memory-Deprived):}
%\textcolor{black}{
    The ablated path removes the contribution of the retrieved
    external-memory representation while preserving the host
    representation. Operationally, the processed memory contribution is
    replaced by a zero vector:
    \begin{equation}
        p_{\text{processed}}^{\text{ablated}}
        =
        \mathbf{0}.
    \end{equation}
\\
    Therefore,
    \begin{equation}
        z_{\text{final}}^{\text{ablated}}
        =
        z
        +
        g\,\mathbf{0}
        =
        z,
    \end{equation}
    and
    \begin{equation}
        y_{\text{ablated}}
        =
        f_{\text{head}}(z).
    \end{equation}
\\
    The ablated output is used to compute the dependency-inducing
    objective described below.
%    }
\end{enumerate}

%\textcolor{black}{
\textbf{Significance of the dual-path design.}
The two paths distinguish prediction using the memory-coupled
representation from prediction using the host representation alone.
During readaptation, the dependency-inducing objective discourages the
host-only path from retaining high confidence in the downstream targets
of support samples associated with the requested identities, while the
normal path is optimized to maintain the corresponding downstream-task
objective under the modified external-memory state.
\\
This mechanism does not imply that identity-related information is
stored exclusively in the external memory. Likewise, prototype removal
alone does not constitute a formal guarantee that all information
associated with the requested identity has been removed from the
host-model parameters.
%}

\begin{figure}[h!]
    \centering
    \includegraphics[width=1\linewidth]{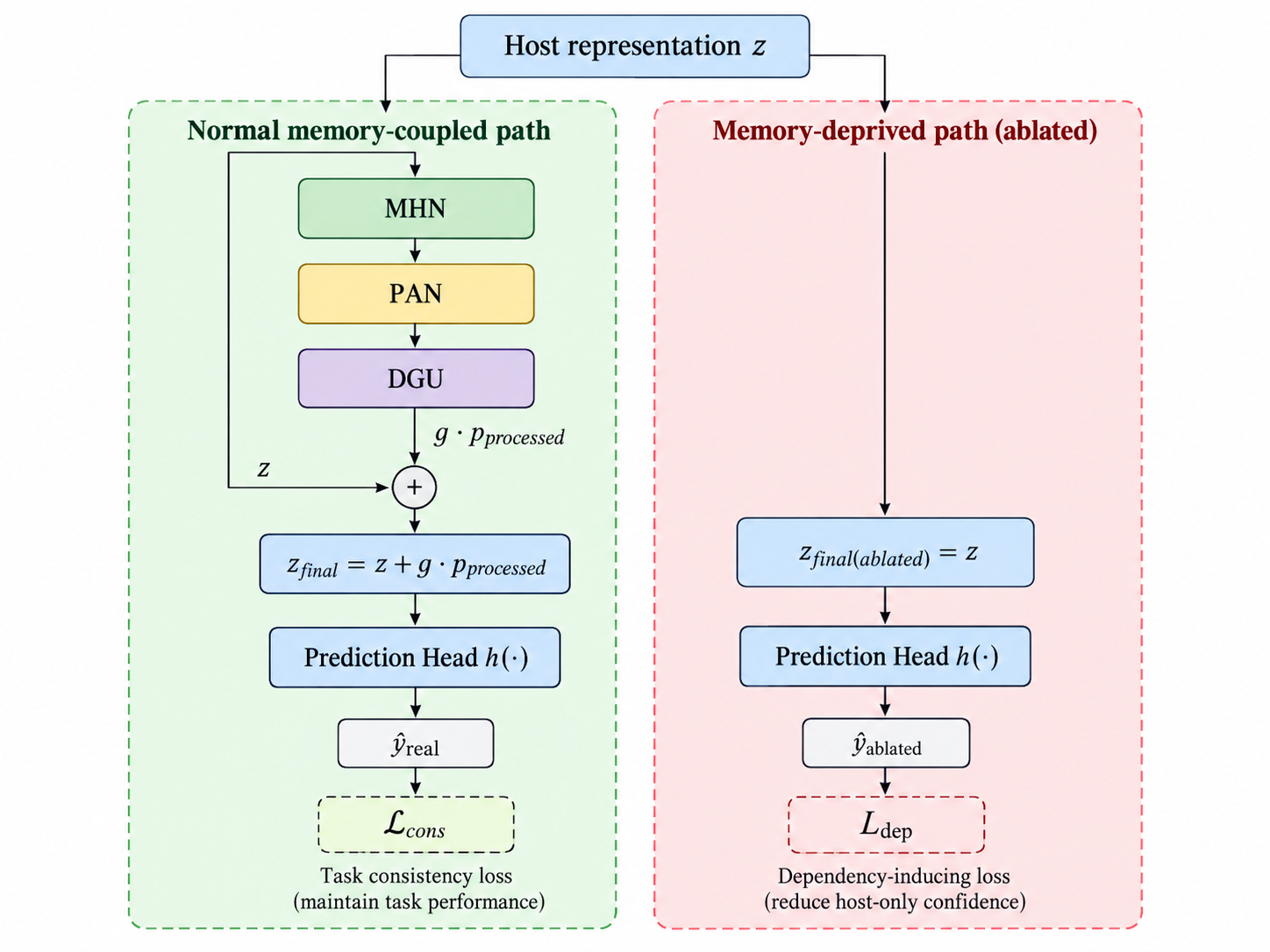}
    \caption{Flowchart of the MCU dual-path forward propagation
    mechanism. The memory-coupled path forms the final representation
    through residual fusion,
    $z_{\mathrm{final}}=z+g\,p_{\mathrm{processed}}$, whereas the
    memory-deprived path removes the external-memory contribution and
    therefore reduces to the host representation $z$.}
    \label{fig:dual-path-forward-flowchart}
\end{figure}

\subsection{Two-Stage Unlearning Process}
\label{subsec:unlearning-process}

\begin{figure}[h!]
    \centering
    \includegraphics[width=1\linewidth]{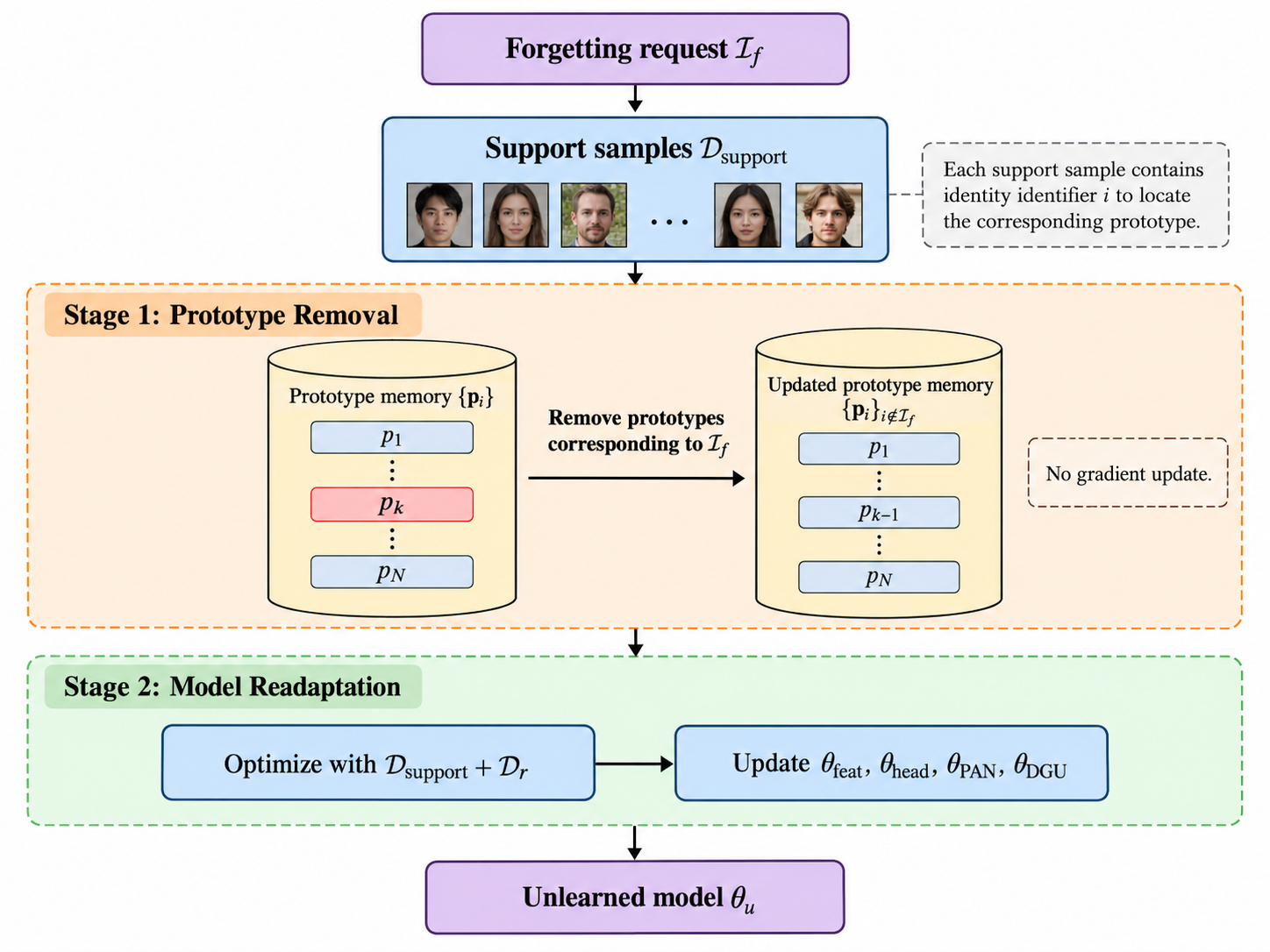}
    \caption{Flowchart of the MCU two-stage unlearning process.
    Stage~1 explicitly removes the prototypes associated with the
    requested identities. Stage~2 performs gradient-based readaptation
    using the support and retain sets under the modified prototype
    memory.}
    \label{fig:unlearning-process-flowchart}
\end{figure}

%\textcolor{black}{
MCU processes a forgetting request through two stages: explicit
prototype removal followed by gradient-based model readaptation
(Figure~\ref{fig:unlearning-process-flowchart}).
%}

\subsubsection{Stage 1: Memory Manipulation --- Prototype Removal}
\label{subsubsec:stage1}

%\textcolor{black}{
Given a forgetting request for the identities in
$\mathcal{I}_{f}$, MCU uses the associated identity identifiers in
$\mathcal{D}_{\text{support}}$ to locate the corresponding entries in
the prototype memory. These target prototypes are then explicitly
removed from the MHN memory bank.
\\
This stage modifies only the external memory and does not itself
perform gradient-based optimization of the host model. After removal,
the target prototypes no longer participate in subsequent MHN
association, which is performed over the remaining stored prototypes.
%}

\subsubsection{Stage 2: Model Readaptation --- Guided Fine-Tuning}
\label{subsubsec:stage2}
%\textcolor{black}{
Prototype removal alone does not guarantee that the influence
associated with the requested identities has been eliminated from the
host-model parameters. MCU therefore performs a subsequent
gradient-based readaptation stage using both the support set
$\mathcal{D}_{\text{support}}$ and the retain set
$\mathcal{D}_{r}$.
\\
Importantly, support samples are still passed through the normal
memory-coupled forward path after target-prototype removal. Because the
target prototype has already been removed, the MHN association for
these samples is performed only over the remaining prototype memory.
The same support samples are also evaluated through the memory-deprived
path to construct the dependency-inducing objective.
\\
For a support sample $(x_s,y_s)$, let
$y_{\text{real}}^{s}$ denote the normal memory-coupled prediction and
$y_{\text{ablated}}^{s}$ denote the corresponding memory-deprived
prediction. For a retained sample $(x_r,y_r)$, let
$y_{\text{real}}^{r}$ denote the normal memory-coupled prediction.
\\
The total readaptation objective is
\begin{equation}
    L_{\text{total}}
    =
    L_{\text{cons}}
    +
    \lambda_{\text{dep}}L_{\text{dep}}
    +
    \lambda_{\text{retain}}L_{\text{retain}}.
\end{equation}
The task-consistency loss is computed on the normal path for the
support samples:
\begin{equation}
    L_{\text{cons}}
    =
    \ell_{\text{task}}
    \left(
        y_{\text{real}}^{s},
        y_s
    \right).
\end{equation}
Minimizing $L_{\text{cons}}$ encourages the normal memory-coupled path
to preserve the supervised downstream-task objective during
readaptation. It should therefore not be interpreted as directly
forcing the model away from the correct downstream prediction.
\\
To preserve utility on retained data, MCU additionally minimizes
\begin{equation}
    L_{\text{retain}}
    =
    \ell_{\text{task}}
    \left(
        y_{\text{real}}^{r},
        y_r
    \right).
\end{equation}
Here, $\ell_{\text{task}}$ denotes the task-specific supervised loss:
cross-entropy for single-label classification and binary
cross-entropy for multi-label prediction.
%}
\\
%\textcolor{black}{
The dependency-inducing loss $L_{\text{dep}}$ is computed using the
memory-deprived output of the support samples. Its specific form
depends on the downstream task.
\\
For single-label classification, MCU minimizes the predicted
probability assigned by the memory-deprived path to the ground-truth
class:
\begin{equation}
    L_{\text{dep}}
    =
    \mathbb{E}
    \left[
        \operatorname{softmax}
        \left(
            y_{\text{ablated}}^{s}
        \right)_{y_s}
    \right].
\end{equation}
\\
For multi-label prediction, let
$\mathcal{P}(y_s)$ denote the set of positive target labels for a
support sample. The dependency-inducing loss is defined as
\begin{equation}
    L_{\text{dep}}
    =
    \frac{1}
    {|\mathcal{P}(y_s)|}
    \sum_{k\in\mathcal{P}(y_s)}
    \sigma
    \left(
        y_{\text{ablated},k}^{s}
    \right),
\end{equation}
when at least one positive target label is present. If no positive
target label is present, this term contributes zero for that sample.
\\
By minimizing $L_{\text{dep}}$, MCU reduces the host-only confidence
on the downstream targets associated with support samples of the
requested identities. This objective should not be interpreted as a
formal proof that identity information has been completely removed;
rather, it is an optimization mechanism used during model readaptation.
\\
The combined objective therefore has three distinct roles:
$L_{\text{cons}}$ preserves task consistency for the support samples
under the modified memory state,
$L_{\text{dep}}$ regulates the memory-deprived path, and
$L_{\text{retain}}$ preserves downstream-task utility on retained data.
%}

% --- Required packages in the preamble ---
% \usepackage{algorithm}
% \usepackage{algpseudocode}
% -----------------------------------------

\begin{algorithm}[ht]
\caption{Memory-Coupled Unlearning (MCU)}
\label{alg:mcu-unlearning}
\begin{algorithmic}[1]

    \Require
        $\theta$: Parameters of the trained host model and
        memory-interaction modules;
        $H$: External MHN memory containing identity-associated prototypes;
        $\mathcal{D}_{\text{support}}$: Support set associated with the
        identities to be forgotten;
        $\mathcal{D}_{r}$: Retain set;
        $\eta$: Learning rate;
        $\lambda_{\text{dep}}$: Weight of the dependency-inducing loss;
        $\lambda_{\text{retain}}$: Weight of the retain-utility loss;
        $E$: Number of readaptation epochs.

    \Ensure
        $\theta_u$: Readapted model parameters;
        $H'$: External memory after target-prototype removal.

    \Statex
    \Statex \textbf{Stage 1: Memory Manipulation --- Prototype Removal}

    \State $\mathcal{I}_{f}
        \gets
        \text{GetIdentitiesFrom}(\mathcal{D}_{\text{support}})$

    \State $H'
        \gets
        \text{RemovePrototypes}(H,\mathcal{I}_{f})$
        \Comment{Remove target identity prototypes.}

    \Statex
    \Statex \textbf{Stage 2: Model Readaptation --- Guided Fine-Tuning}

    \State $\theta_u \gets \theta$

    \State $\Theta_{\text{update}}
        \gets
        \{
        \theta_{\text{feat}},
        \theta_{\text{head}},
        \theta_{\text{PAN}},
        \theta_{\text{DGU}}
        \}$
        \Comment{MHN prototypes remain fixed.}

    \State $\text{Optimizer}
        \gets
        \text{AdamW}(\Theta_{\text{update}},\text{lr}=\eta)$

    \For{$e = 1$ \textbf{to} $E$}

        \For{each readaptation step using available minibatches from
        $\mathcal{D}_{\text{support}}$ and $\mathcal{D}_{r}$}

            \State $L_{\text{total}} \gets 0$

            \If{a support minibatch $(x_s,y_s)$ is available}

                \State $y_{\text{real}}^{s},y_{\text{ablated}}^{s}
                    \gets
                    f_{\theta_u}
                    (x_s,H',
                    \text{compute\_ablated}=\text{True})$

                \State $L_{\text{cons}}
                    \gets
                    \ell_{\text{task}}
                    (y_{\text{real}}^{s},y_s)$
                    \Comment{Preserve downstream-task consistency.}

                \State $L_{\text{dep}}
                    \gets
                    \text{DependencyLoss}
                    (y_{\text{ablated}}^{s},y_s)$
                    \Comment{Reduce host-only target confidence.}

                \State $L_{\text{total}}
                    \gets
                    L_{\text{total}}
                    +
                    L_{\text{cons}}
                    +
                    \lambda_{\text{dep}}L_{\text{dep}}$

            \EndIf

            \If{a retain minibatch $(x_r,y_r)$ is available}

                \State $y_{\text{real}}^{r}
                    \gets
                    f_{\theta_u}
                    (x_r,H',
                    \text{compute\_ablated}=\text{False})$

                \State $L_{\text{retain}}
                    \gets
                    \ell_{\text{task}}
                    (y_{\text{real}}^{r},y_r)$
                    \Comment{Preserve utility on retained data.}

                \State $L_{\text{total}}
                    \gets
                    L_{\text{total}}
                    +
                    \lambda_{\text{retain}}
                    L_{\text{retain}}$

            \EndIf

            \State Update $\Theta_{\text{update}}$ using
                $\nabla_{\Theta_{\text{update}}}L_{\text{total}}$

        \EndFor
    \EndFor

    \State \Return $\theta_u,H'$

\end{algorithmic}
\end{algorithm}
%\textcolor{black}{
In Algorithm~\ref{alg:mcu-unlearning},
$\ell_{\text{task}}$ denotes the task-specific supervised loss:
cross-entropy for single-label classification and binary cross-entropy
for multi-label prediction. The function
$\text{DependencyLoss}(\cdot)$ corresponds to the dependency-inducing
loss defined in Section~\ref{subsec:unlearning-process}. Specifically,
it minimizes the ground-truth softmax probability for single-label
classification and the sigmoid probabilities of positive target labels
for multi-label prediction.
%}
\section{Threat Model and Empirical Security Objective}
\label{sec:threat_model}
\textcolor{black}{
To clarify the scope of the privacy evaluation, we define the attacker
capabilities, defender knowledge, and empirical security objective
considered in this work. The current evaluation focuses on
black-box membership inference and does not claim a formal guarantee
of complete identity-information removal.
}
\textcolor{black}{
\subsection{Defender Knowledge and Data Availability}
The defender is the model provider responsible for processing an
identity-level forgetting request for a set of target identities
$\mathcal{I}_{f}$. At request time, the defender has access to the
trained MCU model, the external prototype memory, the support set
$\mathcal{D}_{\text{support}}$ associated with the requested
identities, and the retain set $\mathcal{D}_{r}$.
The complete original forget set $\mathcal{D}_{f}$ is not directly
revisited during gradient-based model readaptation. Instead, the
identity identifiers associated with
$\mathcal{D}_{\text{support}}$ are used to locate and remove the
corresponding prototype entries from the external memory, while
$\mathcal{D}_{\text{support}}$ and $\mathcal{D}_{r}$ are used during
the subsequent readaptation stage.
The external prototype memory is constructed before request-time
unlearning and is therefore treated as part of the MCU system available
to the defender. As discussed in
Section~\ref{subsec:constructing-memory}, these prototypes are
privacy-sensitive derived representations rather than raw training
samples.
\subsection{Attacker's Goal}
The adversary attempts to determine whether samples associated with
the requested identities still exhibit membership characteristics of
the original training data after unlearning. In the present study, this
goal is evaluated using a MIA~\cite{shokri2017membership, zarifzadeh2023low}.
Specifically, the attacker seeks to distinguish whether a queried
sample is more likely to have originated from the original training
distribution as a member or from the corresponding non-member
distribution. Successful membership inference therefore indicates that
the unlearned model still exposes distinguishable membership-related
behavior for the evaluated samples.
\subsection{Attacker's Capabilities}
We consider a \textbf{black-box adversary} with query access to the
final unlearned model $f_{\theta_u}$. The adversary can submit arbitrary
inputs and observe the corresponding model outputs, such as prediction
scores or probabilities.
\\
The adversary is assumed to know the general unlearning procedure but
does not have direct access to the internal model parameters or the
stored prototype representations. The present evaluation does not
consider an adaptive adversary that explicitly exploits the internal
structure of the MHN, PAN, DGU, or the prototype-removal mechanism.
\subsection{Empirical Security Objective}
The empirical privacy objective of MCU is to reduce the membership
distinguishability of samples associated with the requested identities
after unlearning, while preserving downstream-task utility on retained
data.
\\
Let $A_{\text{MIA}}$ denote the AUC of the evaluated membership
inference attack. An AUC close to $0.5$ indicates that the evaluated
attack cannot reliably distinguish members from non-members under the
considered setting. Importantly, an AUC substantially above or below
$0.5$ may both indicate membership distinguishability depending on the
orientation of the attack score. Therefore, privacy is interpreted in
terms of the deviation from random guessing, i.e.,
$|A_{\text{MIA}}-0.5|$, rather than by assuming that a smaller AUC is
always preferable.
\\
A model retrained from scratch without the requested identities,
$f_{\theta_r}$, is used as an \textbf{empirical reference} for
comparison. However, the present study does not claim formal or
statistical indistinguishability between $f_{\theta_u}$ and
$f_{\theta_r}$. Likewise, an MIA AUC close to $0.5$ should be
interpreted as reduced membership distinguishability under the
evaluated attack rather than as evidence of complete information
removal.
\subsection{Scope of the Privacy Evaluation}
The current empirical privacy evaluation is limited to black-box
membership inference. Adaptive MCU-aware membership inference,
prototype leakage, prototype inversion, model inversion, and
data-reconstruction attacks are not empirically evaluated in the
present study.
Accordingly, these threats are outside the empirically validated scope
of the current work. Evaluating such attacks remains an important
direction for future work and is necessary for a broader assessment of
the privacy properties of external-memory-based unlearning.
}
\section{Experiments}
\label{ch:experiments}

\subsection{Benchmark, Datasets, and Setup}

\textbf{Benchmark.}
The experimental data partitioning and one-shot support construction
are based on the \textbf{1-SHUI (One-Shot Unlearning of Personal
Identities) benchmark}~\cite{de2025unlearning}. Specifically, we adopt
its identity-level partitioning strategy and construct one held-out
support sample for each identity selected for forgetting.

Our MCU implementation differs from a strictly support-only setting in
that the request-time model-readaptation stage additionally uses the
retain set $\mathcal{D}_{r}$ to preserve downstream-task utility.
Therefore, the experimental setting should be interpreted as adopting
the identity partitioning and one-shot support construction of 1-SHUI,
rather than as a strict support-only implementation of the original
benchmark.

\textbf{Datasets.}
Experiments were conducted on three public datasets widely used in
face analysis:
\begin{itemize}
    \item \textbf{CelebA-HQ}~\cite{karras2017progressive}:
    A high-quality ($1024\times1024$) facial-image dataset with
    40 binary facial attributes. Its high image quality and relatively
    limited number of samples per identity provide a challenging
    setting for identity-level unlearning with limited target-specific
    support information.

    \item \textbf{CelebA}~\cite{liu2015deep}:
    A large-scale facial-image dataset containing more than 200,000
    images from over 10,000 identities. The downstream task considered
    in our experiments is multi-label facial-attribute prediction.

    \item \textbf{MUFAC}~\cite{choi2023towards}:
    A facial-image dataset collected from family albums and containing
    substantial age variation within the same identities. The
    downstream task considered in our experiments is age-group
    classification.
\end{itemize}

Although the evaluated datasets contain facial images and identity
annotations, the downstream tasks considered in this work should not
be interpreted as an operational biometric face-verification setting.
Identity defines the unit to which a forgetting request applies,
whereas facial attributes or age groups define the downstream
prediction targets. Accordingly, the experiments evaluate
identity-level unlearning on facial datasets rather than biometric
verification based on enrollment templates, embedding similarity, and
decision thresholds.

\begin{figure}[h!]
    \centering
    \includegraphics[width=0.9\linewidth]{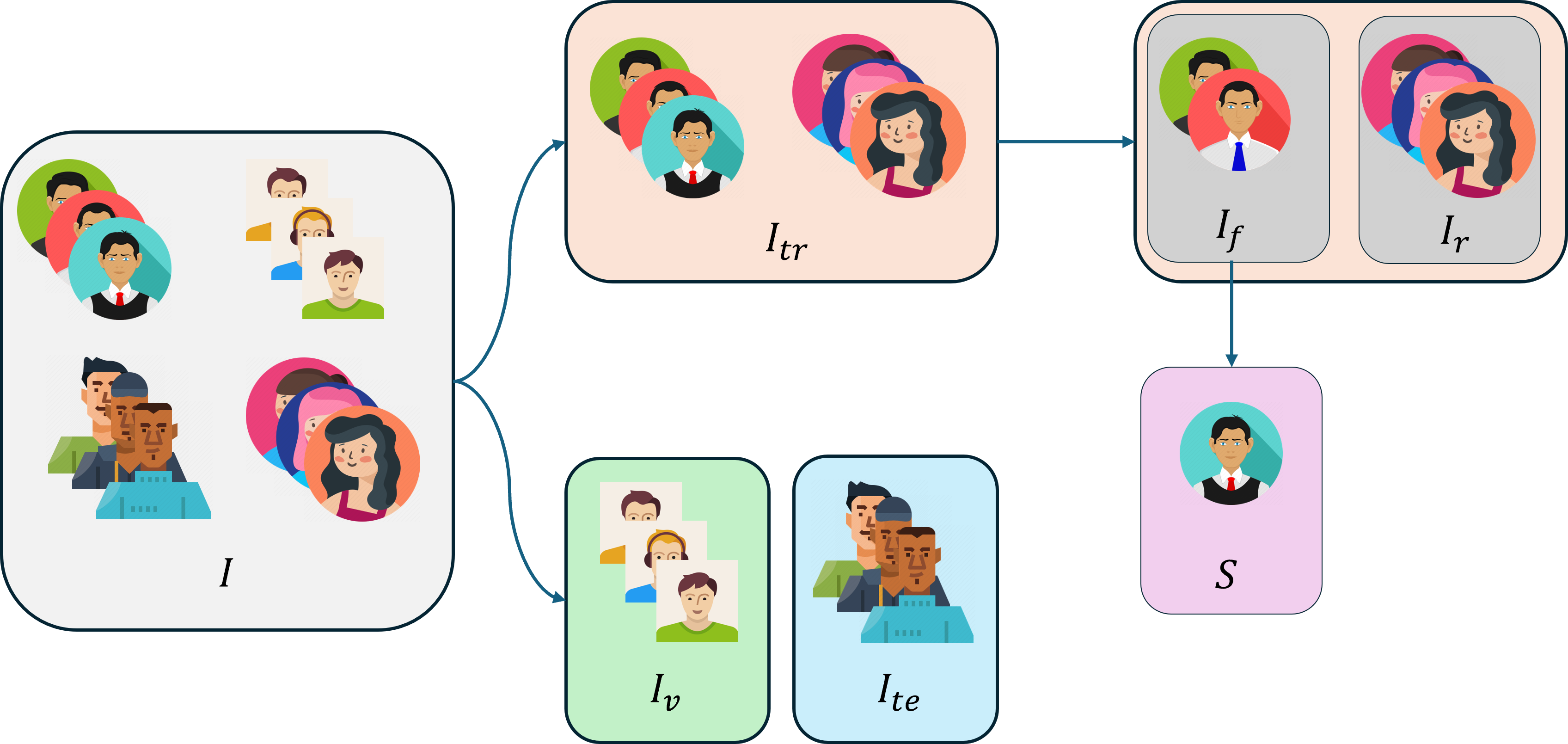}
    \caption{Identity partitioning and one-shot support construction
    adopted from the 1-SHUI benchmark. During request-time
    readaptation, MCU additionally uses the retain set
    $\mathcal{D}_{r}$.}
    \label{fig:data-construction}
\end{figure}

The identity partitioning follows the procedure of
De Min et al.~\cite{de2025unlearning}, as illustrated in
Figure~\ref{fig:data-construction}. The procedure consists of the
following steps:

\begin{enumerate}
    \item \textbf{First-Level Split: Training, Validation, and Test
    Identities.}
    All identities are partitioned into mutually exclusive training,
    validation, and test identity sets, denoted by
    $\mathcal{I}_{tr}$, $\mathcal{I}_{v}$, and
    $\mathcal{I}_{te}$, respectively.

    \item \textbf{Second-Level Split: Forget and Retain Identities.}
    A specified number of identities are sampled from
    $\mathcal{I}_{tr}$ to form the forget identity set
    $\mathcal{I}_{f}$. The remaining training identities constitute
    the retain identity set $\mathcal{I}_{r}$.

    \item \textbf{Third-Level Split: One-Shot Support Construction.}
    For each identity in $\mathcal{I}_{f}$, one image is selected as
    a held-out support sample. These samples collectively form
    $\mathcal{D}_{\text{support}}$. Following the benchmark
    partitioning procedure, the designated support samples are excluded
    from host-model pretraining.
\end{enumerate}

The host model is pretrained using samples associated with
$\mathcal{I}_{f}$ and $\mathcal{I}_{r}$ according to the benchmark
partitioning, with the designated support samples excluded from
host-model pretraining. The external prototype memory is subsequently
constructed during the MCU preparation stage as described in
Section~\ref{subsec:constructing-memory}.

At request time, the complete original forget set
$\mathcal{D}_{f}$ is not directly revisited for gradient-based
readaptation. Instead, MCU performs model readaptation using the
one-shot support set $\mathcal{D}_{\text{support}}$ together with the
retain set $\mathcal{D}_{r}$.

\textbf{Baseline Methods.}
We compare MCU against several representative unlearning approaches
and reference models:
\begin{itemize}
    \item \textbf{Reference Models:}
    \textit{Pretrain}, representing the original model before
    unlearning, and \textit{Retrain}, representing a model retrained
    from scratch after excluding the identities requested for
    forgetting.

    \item \textbf{Compared Unlearning Baselines:}
    \textit{SSD}~\cite{foster2024fast},
    \textit{PGU}~\cite{hoang2024learn},
    \textit{JiT}~\cite{foster2024zero}, and
    \textit{MetaUnlearn}~\cite{de2025unlearning}.

    \item \textbf{Full-Data Reference Methods:}
    \textit{SCRUB}~\cite{kurmanji2023towards} and
    \textit{Bad Teacher}~\cite{chundawat2023can}, which require broader
    access to training-related data during unlearning.
\end{itemize}

The evaluated approaches do not necessarily operate under identical
request-time data-access assumptions. Therefore, the comparisons are
interpreted as empirical performance comparisons rather than as
evidence that all methods use exactly the same information during
unlearning. In particular, MCU uses both
$\mathcal{D}_{\text{support}}$ and $\mathcal{D}_{r}$ during
request-time model readaptation.

\textbf{Implementation Details.}
All experiments were conducted on a single NVIDIA GeForce RTX 3080 GPU
using PyTorch. The host model for all experiments is a ResNet-18
architecture. Unless otherwise specified, MCU uses one readaptation
epoch ($E=1$), a learning rate of $1\times10^{-4}$,
a dependency-inducing loss weight of
$\lambda_{\text{dep}}=1.0$, and a retain-utility loss weight of
$\lambda_{\text{retain}}=1.0$.

During request-time model readaptation, the feature extractor,
prediction head, PAN, and DGU are jointly optimized using AdamW.
The remaining MHN prototype patterns are kept fixed and are not updated
through backpropagation after the target prototypes have been removed.

\subsection{Evaluation Metrics}

The evaluation considers three complementary dimensions:
\textit{downstream-task utility},
\textit{membership distinguishability}, and
\textit{computational efficiency}.

%\textcolor{black}{
\textbf{Model Utility.}
We evaluate downstream-task performance on the retain set
$\mathcal{D}_{r}$, forget set $\mathcal{D}_{f}$, and test set
$\mathcal{D}_{te}$. Following the adopted experimental protocol,
Mean Average Precision (mAP) is used as the reported downstream-task
utility metric for CelebA-HQ, CelebA, and MUFAC, covering the
facial-attribute prediction tasks on CelebA-HQ and CelebA and the
age-group classification task on MUFAC.
We additionally report the \textbf{Tug of War (ToW)}
metric~\cite{zhao2024makes}, which summarizes the discrepancy between
an unlearned model and the retrained reference model across the
evaluated data partitions.
%}

\textbf{Membership Inference Evaluation.}
We evaluate membership distinguishability using the Relativity-based
\textbf{Membership Inference Attack (MIA)}
~\cite{zarifzadeh2023low}. The primary metric is the
\textbf{Area Under the ROC Curve (AUC)}.

An MIA AUC close to $0.5$ indicates that the evaluated attack performs
close to random guessing under the considered setting. Importantly,
an AUC substantially above or below $0.5$ may both indicate membership
distinguishability depending on the orientation of the attack score.
Therefore, we interpret membership distinguishability primarily in
terms of the deviation
\begin{equation}
    \left|\mathrm{AUC}-0.5\right|.
\end{equation}

We additionally report the \textbf{True Positive Rate (TPR)} at a
\textbf{False Positive Rate (FPR) of $0.1\%$} to characterize the
attack at a low-FPR operating point. This metric is used to examine
high-confidence membership inference and is not intended to represent
an operational biometric decision threshold.

\textbf{Efficiency.}
We report the request-time \textbf{SpeedUp} factor as
\begin{equation}
    \mathrm{SpeedUp}
    =
    \frac{T_{\text{retrain}}}{T_{\text{unlearn}}}.
\end{equation}

For MCU, $T_{\text{unlearn}}$ measures the request-time prototype
removal and gradient-based model-readaptation stages. The one-time
prototype-construction procedure performed during MCU preparation is
not included in $T_{\text{unlearn}}$. Therefore, the reported SpeedUp
represents request-time efficiency after prototype preparation rather
than end-to-end lifecycle speedup.

\textbf{A Complementary Metric: The Utility-Privacy Balance Index
(UPBI).}

Because model utility and membership-inference measurements characterize
different aspects of unlearning performance, we additionally report the
UPBI as a complementary
descriptive measure of their observed trade-off. UPBI is not intended
to serve as a universal privacy metric or to replace the individual
utility and attack-specific privacy measurements.

The UPBI is defined as
\begin{equation}
    \mathrm{UPBI}
    =
    \sqrt{\mathrm{ToW}'\times\mathrm{CPS}},
\end{equation}
where $\mathrm{ToW}'$ denotes the normalized utility score and
$\mathrm{CPS}$ denotes the composite privacy score.

The normalized utility score is defined as
\begin{equation}
    \mathrm{ToW}'
    =
    \epsilon
    +
    (1-\epsilon)
    \times
    \frac{
        \mathrm{ToW}-\min(\mathrm{ToW})
    }{
        \max(\mathrm{ToW})-\min(\mathrm{ToW})
    },
\end{equation}
where $\epsilon=0.01$.

The composite privacy score is defined as
\begin{equation}
    \mathrm{CPS}
    =
    \sqrt{\mathrm{PSS}\times\mathrm{HCRS}},
\end{equation}
where
\begin{equation}
    \mathrm{PSS}
    =
    1-\left|\mathrm{AUC}-0.5\right|,
\end{equation}
and
\begin{equation}
    \mathrm{HCRS}
    =
    1-\mathrm{TPR}.
\end{equation}

Under this definition, PSS is maximized when the evaluated MIA AUC is
$0.5$ and decreases as the AUC moves away from random guessing in
either direction. The geometric mean is used so that a low value in
either the utility or privacy component reduces the resulting UPBI
score.

Because $\mathrm{ToW}'$ is normalized using the methods included in a
particular comparison and depends on the selected parameter
$\epsilon$, UPBI should be interpreted as a descriptive summary under
the evaluated configuration rather than as a method-independent or
universal ranking criterion. Accordingly, all UPBI results are
considered together with the individual ToW, MIA AUC, and TPR
measurements.

\subsection{Main Results and Analysis}

The main experimental results comparing MCU against the evaluated
baselines are presented in
Tables~\ref{tab:celebahq-table},
\ref{tab:celeba-table}, and
\ref{tab:mufac-table}.
Across these tables, bold and dagger ($\dagger$) denote the best
and second-best results, respectively, among the methods in the
limited-target-data comparison group. For MIA AUC, performance is
ranked according to $|\mathrm{AUC}-0.5|$, and TPR is reported at an
FPR of $0.1\%$. The evaluated methods may operate under different
request-time data-access assumptions. In particular, MCU uses
$\mathcal{D}_{\mathrm{support}}$ together with $\mathcal{D}_{r}$
during model readaptation, while the complete original forget set
$\mathcal{D}_{f}$ is not directly revisited.

\begin{table*}[t!]
\centering
\caption
{
Comparison of unlearning methods on CelebA-HQ for
20 and 50 identity-forgetting tasks. Model utility is reported using {\upshape mAP (\%)}.
}
\label{tab:celebahq-table}

\resizebox{\textwidth}{!}{%
\begin{tabular}{l|ccc|c|cc}
\hline
\multirow{2}{*}{\textbf{Method}}
& \multicolumn{3}{c|}{\textbf{mAP (\%) $\uparrow$}}
& \multirow{2}{*}{\textbf{ToW $\uparrow$}}
& \multicolumn{2}{c}{\textbf{MIA}} \\
\cline{2-4} \cline{6-7}
& $\mathcal{D}_r$
& $\mathcal{D}_f$
& $\mathcal{D}_{te}$
&
& \textbf{AUC (ideal$\rightarrow$0.5)}
& \textbf{TPR (\%) $\downarrow$} \\
\hline

\multicolumn{7}{c}{\textit{\textbf{20 Identities}}} \\
\hline

Pretrain
& 70.8$\pm$0.1
& 70.2$\pm$1.1
& 68.9$\pm$0.1
& -
& -
& - \\

Retrain
& 70.7$\pm$0.5
& 65.1$\pm$3.2
& 68.9$\pm$0.6
& -
& -
& - \\

SCRUB~\cite{kurmanji2023towards}
& 71.1$\pm$0.2
& 70.1$\pm$2.2
& 69.9$\pm$0.2
& 93.6$\pm$0.7
& 0.58
& 0.22 \\

Bad Teacher~\cite{chundawat2023can}
& 70.9$\pm$0.0
& 70.8$\pm$1.1
& 69.7$\pm$0.1
& 93.3$\pm$0.1
& 0.59
& 0.15 \\

\hline

SSD~\cite{foster2024fast}
& 68.0$\pm$0.2
& 69.2$\pm$1.0
& 66.8$\pm$0.2
& 90.0$\pm$0.9
& \textbf{0.51}
& \textbf{0.10} \\

PGU~\cite{hoang2024learn}
& 70.83$\pm$0.3
& 70.1$\pm$0.1
& 69.7$\pm$0.1
& 94.1$\pm$0.4$^\dagger$
& 0.57
& 0.22 \\

JiT~\cite{foster2024zero}
& 60.1$\pm$0.1
& 59.0$\pm$0.3
& 59.0$\pm$0.2
& 83.8$\pm$0.2
& 0.58
& 0.13$^\dagger$ \\

MetaUnlearn~\cite{de2025unlearning}
& 59.0$\pm$0.1
& 57.0$\pm$2.0
& 58.7$\pm$0.1
& 72.9$\pm$1.0
& 0.61
& 0.20 \\

\rowcolor{cyan!20}
MCU (Ours)
& 70.7$\pm$0.4
& 69.3$\pm$0.1
& 70.0$\pm$0.0
& \textbf{94.7$\pm$0.5}
& 0.56$^\dagger$
& 0.15 \\

\hline

\multicolumn{7}{c}{\textit{\textbf{50 Identities}}} \\
\hline

Pretrain
& 70.7$\pm$0.1
& 72.2$\pm$0.1
& 68.9$\pm$0.1
& -
& -
& - \\

Retrain
& 70.7$\pm$0.1
& 66.4$\pm$0.3
& 70.6$\pm$0.1
& -
& -
& - \\

SCRUB~\cite{kurmanji2023towards}
& 71.0$\pm$0.1
& 68.2$\pm$0.3
& 69.6$\pm$0.1
& 96.8$\pm$0.3
& 0.55
& 0.16 \\

Bad Teacher~\cite{chundawat2023can}
& 70.7$\pm$0.1
& 68.7$\pm$0.1
& 69.5$\pm$0.1
& 96.6$\pm$0.3
& 0.53
& 0.17 \\

\hline

SSD~\cite{foster2024fast}
& 53.2$\pm$0.3
& 51.7$\pm$0.2
& 51.7$\pm$4.2
& 57.0$\pm$0.2
& \textbf{0.51}
& 0.39$^\dagger$ \\

PGU~\cite{hoang2024learn}
& 70.6$\pm$0.0
& 69.0$\pm$0.1
& 68.7$\pm$0.1
& 94.0$\pm$0.3$^\dagger$
& 0.55$^\dagger$
& 0.39$^\dagger$ \\

JiT~\cite{foster2024zero}
& 48.4$\pm$0.4
& 47.2$\pm$4.4
& 46.7$\pm$0.3
& 47.7$\pm$0.9
& 0.60
& 0.39$^\dagger$ \\

MetaUnlearn~\cite{de2025unlearning}
& 67.4$\pm$0.0
& 66.1$\pm$1.8
& 66.9$\pm$0.2
& 92.8$\pm$0.3
& \textbf{0.51}
& 1.18 \\

\rowcolor{cyan!20}
MCU (Ours)
& 70.8$\pm$0.2
& 68.9$\pm$0.6
& 69.0$\pm$0.3
& \textbf{95.8$\pm$0.3}
& \textbf{0.51}
& \textbf{0.20} \\

\hline
\end{tabular}
}
\end{table*}
\begin{table*}[t!]
\centering
\caption
{
Comparison of unlearning methods on CelebA for
5 and 10 identity-forgetting tasks. Model utility is reported using {\upshape mAP (\%)}.
}
\label{tab:celeba-table}

\resizebox{\textwidth}{!}{%
\begin{tabular}{l|ccc|c|cc}
\hline
\multirow{2}{*}{\textbf{Method}}
& \multicolumn{3}{c|}{\textbf{mAP (\%) $\uparrow$}}
& \multirow{2}{*}{\textbf{ToW $\uparrow$}}
& \multicolumn{2}{c}{\textbf{MIA}} \\
\cline{2-4} \cline{6-7}
& $\mathcal{D}_r$
& $\mathcal{D}_f$
& $\mathcal{D}_{te}$
&
& \textbf{AUC (ideal$\rightarrow$0.5)}
& \textbf{TPR (\%) $\downarrow$} \\
\hline

\multicolumn{7}{c}{\textit{\textbf{5 Identities}}} \\
\hline

Pretrain
& 74.5$\pm$0.0
& 69.3$\pm$0.3
& 73.0$\pm$0.1
& -
& -
& - \\

Retrain
& 74.1$\pm$0.1
& 50.1$\pm$2.1
& 73.2$\pm$0.1
& -
& -
& - \\

SCRUB~\cite{kurmanji2023towards}
& 75.2$\pm$0.1
& 50.6$\pm$1.3
& 70.6$\pm$0.2
& 98.1$\pm$0.0
& 0.54
& 6.7 \\

Bad Teacher~\cite{chundawat2023can}
& 74.1$\pm$0.0
& 49.7$\pm$2.1
& 73.3$\pm$0.1
& 98.3$\pm$0.1
& 0.66
& 13.4 \\

\hline

SSD~\cite{foster2024fast}
& 22.6$\pm$0.1
& 26.1$\pm$1.0
& 22.4$\pm$0.2
& 17.6$\pm$0.3
& 0.21
& 3.0$^\dagger$ \\

PGU~\cite{hoang2024learn}
& 74.3$\pm$0.2
& 50.1$\pm$1.1
& 73.5$\pm$0.1
& \textbf{98.1$\pm$0.5}
& 0.70
& 6.0 \\

JiT~\cite{foster2024zero}
& 71.1$\pm$0.1
& 49.3$\pm$1.5
& 70.4$\pm$0.2
& 92.1$\pm$0.2
& \textbf{0.44}
& 6.0 \\

MetaUnlearn~\cite{de2025unlearning}
& 68.0$\pm$0.5
& 48.4$\pm$1.2
& 67.2$\pm$0.1
& 86.3$\pm$1.0
& 0.40$^\dagger$
& 12.0 \\

\rowcolor{cyan!20}
MCU (Ours)
& 76.3$\pm$0.3
& 50.5$\pm$1.3
& 75.1$\pm$0.2
& 97.5$\pm$0.3$^\dagger$
& 0.40$^\dagger$
& \textbf{1.5} \\

\hline

\multicolumn{7}{c}{\textit{\textbf{10 Identities}}} \\
\hline

Pretrain
& 78.1$\pm$0.1
& 79.5$\pm$1.3
& 76.3$\pm$0.3
& -
& -
& - \\

Retrain
& 84.0$\pm$0.1
& 77.4$\pm$2.0
& 79.7$\pm$0.1
& -
& -
& - \\

SCRUB~\cite{kurmanji2023towards}
& 78.2$\pm$1.2
& 70.3$\pm$2.5
& 77.0$\pm$0.3
& 86.9$\pm$1.0
& 0.52
& 6.3 \\

Bad Teacher~\cite{chundawat2023can}
& 77.9$\pm$0.2
& 71.8$\pm$2.2
& 76.5$\pm$0.1
& 87.5$\pm$1.0
& 0.55
& 11.1 \\

\hline

SSD~\cite{foster2024fast}
& 27.8$\pm$2.5
& 28.6$\pm$3.2
& 27.8$\pm$3.5
& 11.1$\pm$5.0
& 0.37
& 7.4 \\

PGU~\cite{hoang2024learn}
& 78.1$\pm$0.1
& 70.6$\pm$0.5
& 76.6$\pm$0.3
& 86.7$\pm$0.1$^\dagger$
& 0.48$^\dagger$
& 6.3 \\

JiT~\cite{foster2024zero}
& 73.5$\pm$1.2
& 67.4$\pm$2.2
& 72.4$\pm$0.3
& 76.2$\pm$1.3
& \textbf{0.49}
& 3.4$^\dagger$ \\

MetaUnlearn~\cite{de2025unlearning}
& 76.2$\pm$0.2
& 69.3$\pm$2.1
& 74.8$\pm$0.5
& 80.5$\pm$0.0
& 0.52$^\dagger$
& \textbf{3.3} \\

\rowcolor{cyan!20}
MCU (Ours)
& 78.5$\pm$0.1
& 75.0$\pm$0.5
& 77.3$\pm$0.3
& \textbf{89.6$\pm$0.1}
& 0.55
& 5.5 \\

\hline
\end{tabular}
}
\end{table*}
\begin{table*}[t!]
\centering
\caption
{
Comparison of unlearning methods on MUFAC for
5 and 10 identity-forgetting tasks. Model utility is reported using {\upshape mAP (\%)}.
}
\label{tab:mufac-table}

\resizebox{\textwidth}{!}{%
\begin{tabular}{l|ccc|c|cc}
\hline
\multirow{2}{*}{\textbf{Method}}
& \multicolumn{3}{c|}{\textbf{mAP (\%) $\uparrow$}}
& \multirow{2}{*}{\textbf{ToW $\uparrow$}}
& \multicolumn{2}{c}{\textbf{MIA}} \\
\cline{2-4} \cline{6-7}
& $\mathcal{D}_r$
& $\mathcal{D}_f$
& $\mathcal{D}_{te}$
&
& \textbf{AUC (ideal$\rightarrow$0.5)}
& \textbf{TPR (\%) $\downarrow$} \\
\hline

\multicolumn{7}{c}{\textit{\textbf{5 Identities}}} \\
\hline

Pretrain
& 89.0$\pm$0.1
& 86.2$\pm$0.0
& 75.9$\pm$0.1
& -
& -
& - \\

Retrain
& 90.3$\pm$0.0
& 50.9$\pm$0.2
& 75.1$\pm$0.1
& -
& -
& - \\

SCRUB~\cite{kurmanji2023towards}
& 90.1$\pm$0.5
& 81.5$\pm$0.1
& 73.0$\pm$0.2
& 67.8$\pm$0.7
& 0.79
& 3.42 \\

Bad Teacher~\cite{chundawat2023can}
& 89.6$\pm$0.0
& 82.3$\pm$0.3
& 70.7$\pm$0.1
& 65.1$\pm$0.5
& 0.70
& 7.84 \\

\hline

SSD~\cite{foster2024fast}
& 13.7$\pm$0.2
& 12.9$\pm$2.0
& 12.5$\pm$0.1
& 5.4$\pm$0.9
& 0.54$^\dagger$
& 0.11$^\dagger$ \\

PGU~\cite{hoang2024learn}
& 89.4$\pm$0.0
& 83.7$\pm$0.3
& 75.0$\pm$0.1
& 67.0$\pm$0.5$^\dagger$
& 0.76
& 0.86 \\

JiT~\cite{foster2024zero}
& 83.5$\pm$0.1
& 73.3$\pm$0.2
& 65.7$\pm$0.2
& 65.2$\pm$0.2
& 0.68
& 0.77 \\

MetaUnlearn~\cite{de2025unlearning}
& 90.4$\pm$0.0
& 83.5$\pm$0.2
& 70.6$\pm$0.5
& 58.9$\pm$1.0
& 0.75
& 1.28 \\

\rowcolor{cyan!20}
MCU (Ours)
& 90.1$\pm$0.6
& 82.0$\pm$0.3
& 75.5$\pm$0.3
& \textbf{67.8$\pm$0.6}
& \textbf{0.52}
& \textbf{0.10} \\

\hline

\multicolumn{7}{c}{\textit{\textbf{10 Identities}}} \\
\hline

Pretrain
& 96.5$\pm$0.1
& 95.6$\pm$0.1
& 72.2$\pm$0.5
& -
& -
& - \\

Retrain
& 90.3$\pm$0.1
& 54.3$\pm$0.5
& 70.8$\pm$0.1
& -
& -
& - \\

SCRUB~\cite{kurmanji2023towards}
& 85.7$\pm$0.1
& 77.1$\pm$0.7
& 69.4$\pm$0.1
& 75.7$\pm$0.3
& 0.61
& 1.09 \\

Bad Teacher~\cite{chundawat2023can}
& 86.13$\pm$0.1
& 78.2$\pm$0.2
& 69.9$\pm$0.2
& 74.4$\pm$0.0
& 0.00
& 0.00 \\

\hline

SSD~\cite{foster2024fast}
& 21.1$\pm$0.3
& 18.4$\pm$1.2
& 19.7$\pm$0.5
& 6.4$\pm$1.2
& 0.55$^\dagger$
& 0.14$^\dagger$ \\

PGU~\cite{hoang2024learn}
& 86.4$\pm$0.0
& 78.2$\pm$0.3
& 69.9$\pm$0.1
& \textbf{75.7$\pm$0.5}
& 0.66
& 0.70 \\

JiT~\cite{foster2024zero}
& 77.8$\pm$0.4
& 73.9$\pm$0.4
& 65.3$\pm$0.3
& 44.0$\pm$0.9
& 0.61
& 1.09 \\

MetaUnlearn~\cite{de2025unlearning}
& 86.4$\pm$0.0
& 78.2$\pm$1.3
& 69.6$\pm$0.2
& 74.4$\pm$0.3$^\dagger$
& 0.64
& 1.09 \\

\rowcolor{cyan!20}
MCU (Ours)
& 86.4$\pm$0.2
& 80.4$\pm$0.1
& 69.7$\pm$0.3
& 72.7$\pm$0.3
& \textbf{0.53}
& \textbf{0.12} \\

\hline
\end{tabular}
}
\end{table*}

Across the evaluated settings, MCU demonstrates a favorable empirical
trade-off between downstream-task utility and membership
distinguishability. On CelebA-HQ/50, MCU achieves a ToW score of 95.8
with an MIA AUC of 0.51, which is close to random guessing under the
evaluated Relativity-based attack. On MUFAC, MCU obtains MIA AUC values
of approximately 0.52--0.53 while maintaining competitive
downstream-task utility. These results indicate low membership
distinguishability under the evaluated attack, but should not be
interpreted as evidence of complete identity-information removal or as
a general privacy guarantee.

\begin{figure}[ht!]
    \centering
    \includegraphics[width=\linewidth]
    {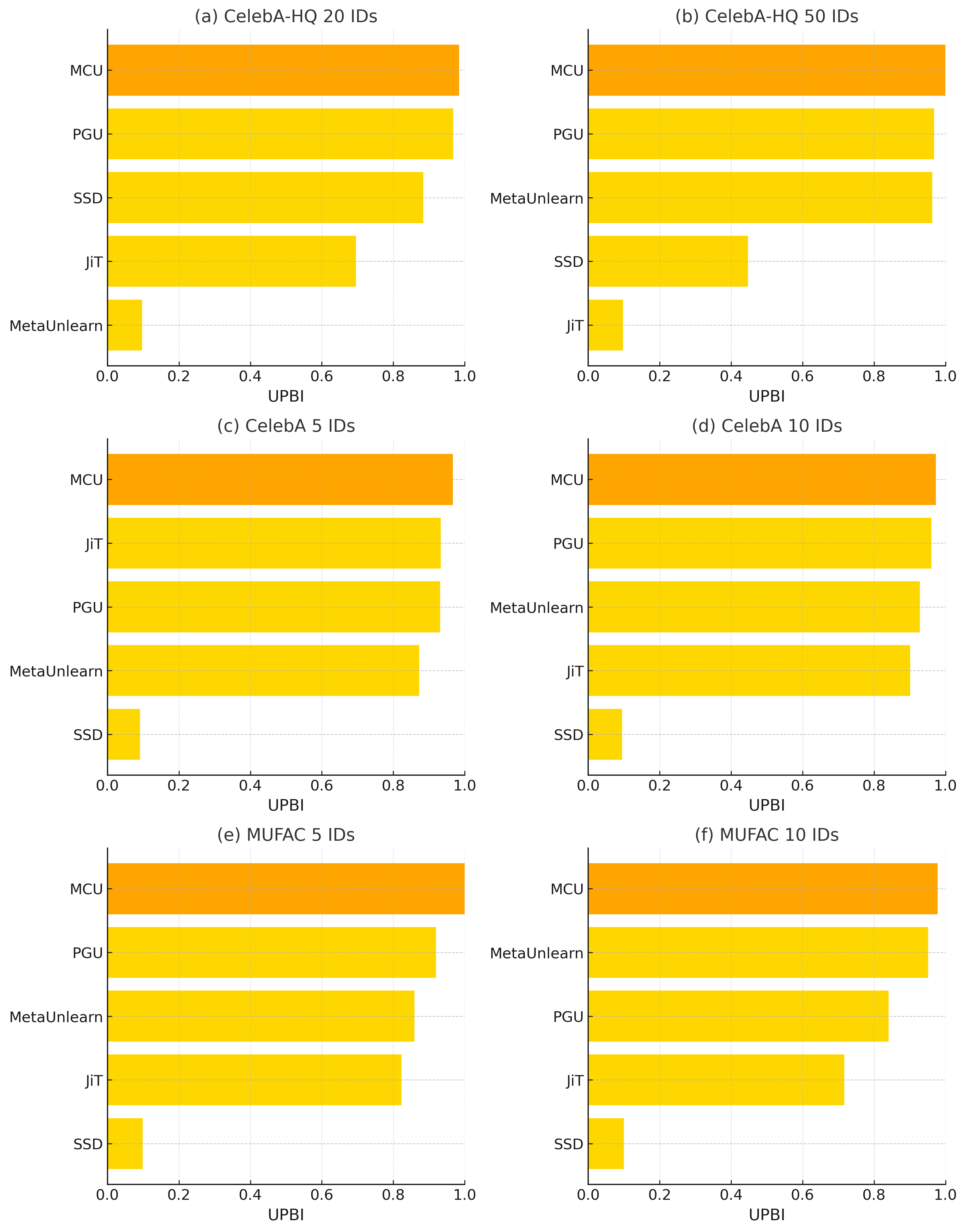}
    \caption{Utility--membership trade-off summarized by UPBI.
    UPBI should be interpreted together with the individual utility
    and membership-inference metrics.}
    \label{fig:upbi_results}
\end{figure}

Figure~\ref{fig:upbi_results} summarizes the observed
utility--membership trade-off using UPBI. Under the current definition
and evaluated method set, MCU obtains the highest UPBI score across the
reported configurations. Because UPBI depends on comparison-dependent
normalization and attack-specific privacy measurements, it is used only
as a descriptive summary rather than as evidence of universal
superiority.

\subsection{Computational Efficiency and Parameter Sensitivity}

\begin{figure}[h!]
    \centering
    \includegraphics[width=0.8\linewidth]{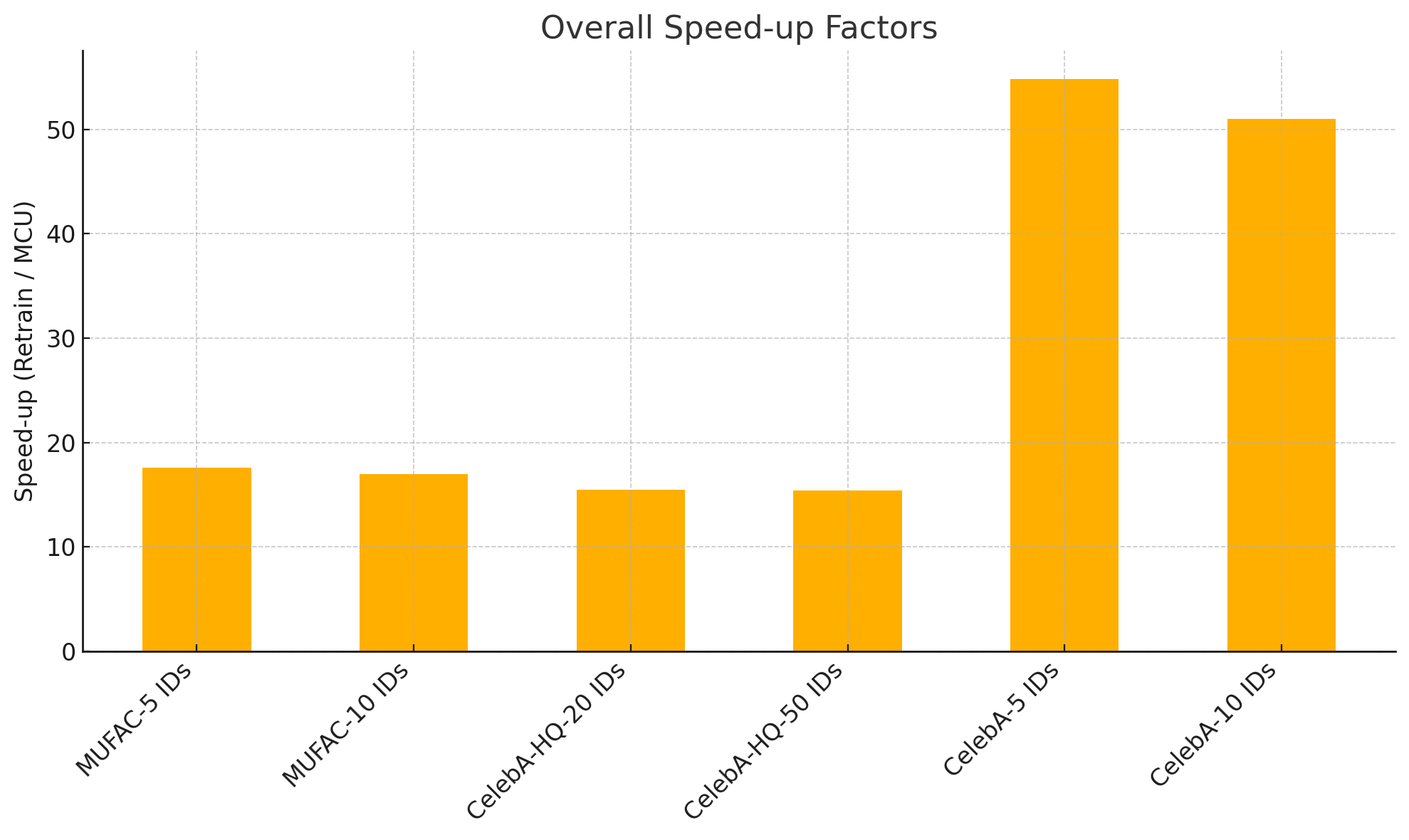}
    \caption{Request-time SpeedUp factors of MCU relative to
    retraining. The one-time prototype-construction cost is not included
    in $T_{\text{unlearn}}$.}
    \label{fig:mcu_vs_retrain_speedup}
\end{figure}

\textbf{Efficiency.}
As shown in Figure~\ref{fig:mcu_vs_retrain_speedup}, MCU achieves up
to a 54.8-fold request-time speedup relative to retraining under the
evaluated settings. As defined above, this runtime includes
target-prototype removal and model readaptation after the external
prototype memory has already been constructed. Therefore, the reported
speedup does not include the one-time prototype-construction cost.

\begin{figure}[h!]
    \centering
    \includegraphics[width=\linewidth]{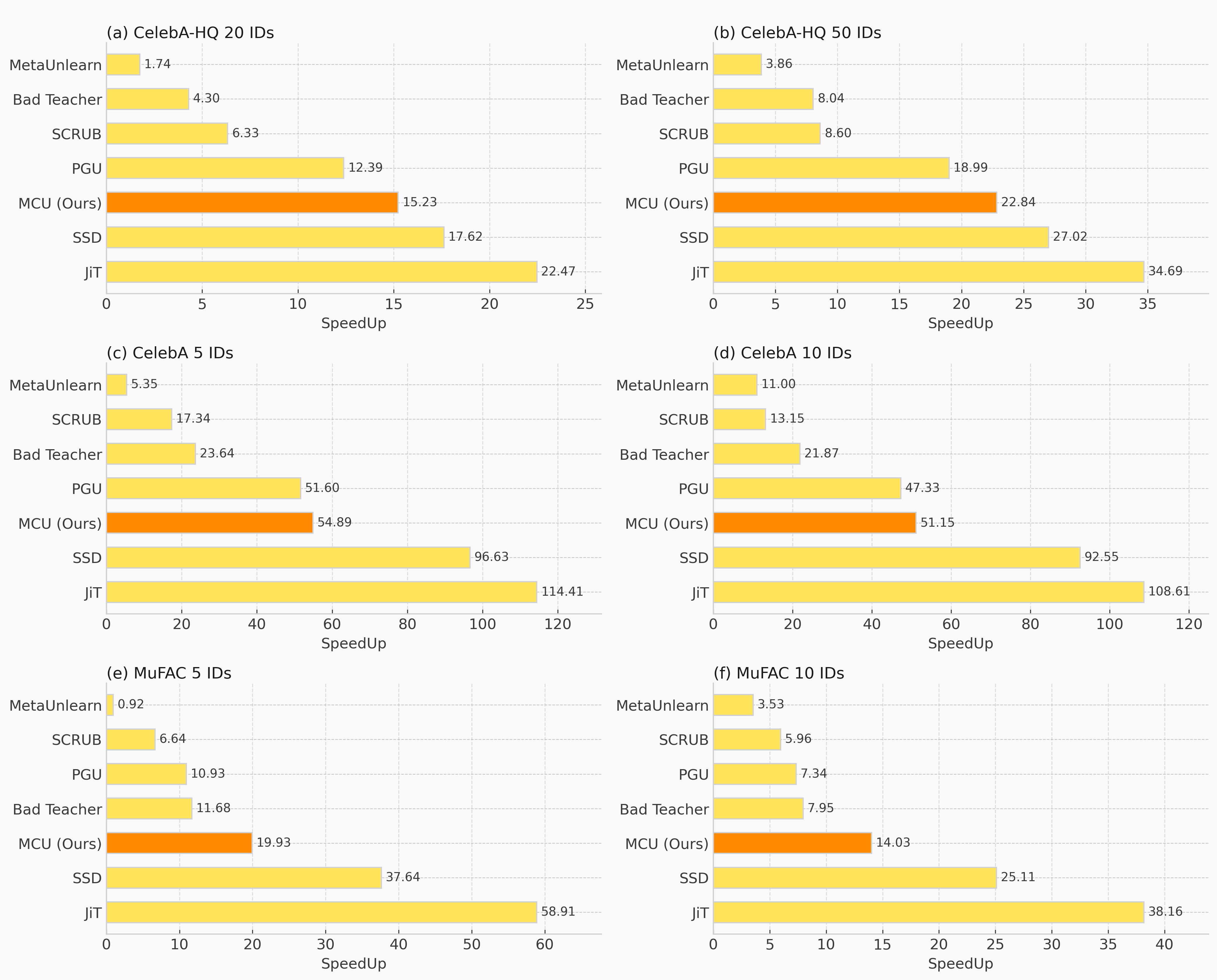}
    \caption{
        \textbf{Request-time SpeedUp factors across the evaluated
        unlearning methods.}
    }
    \label{fig:all_methods_speedup}
\end{figure}

Figure~\ref{fig:all_methods_speedup} further compares request-time
efficiency across the evaluated unlearning methods. Some methods require
less request-time computation than MCU, but the evaluated approaches
exhibit different trade-offs among computational cost,
downstream-task utility, and membership distinguishability.
Accordingly, runtime is considered together with the corresponding
utility and membership-inference measurements.

\begin{figure}[h!]
    \centering
    \includegraphics[width=0.8\linewidth]{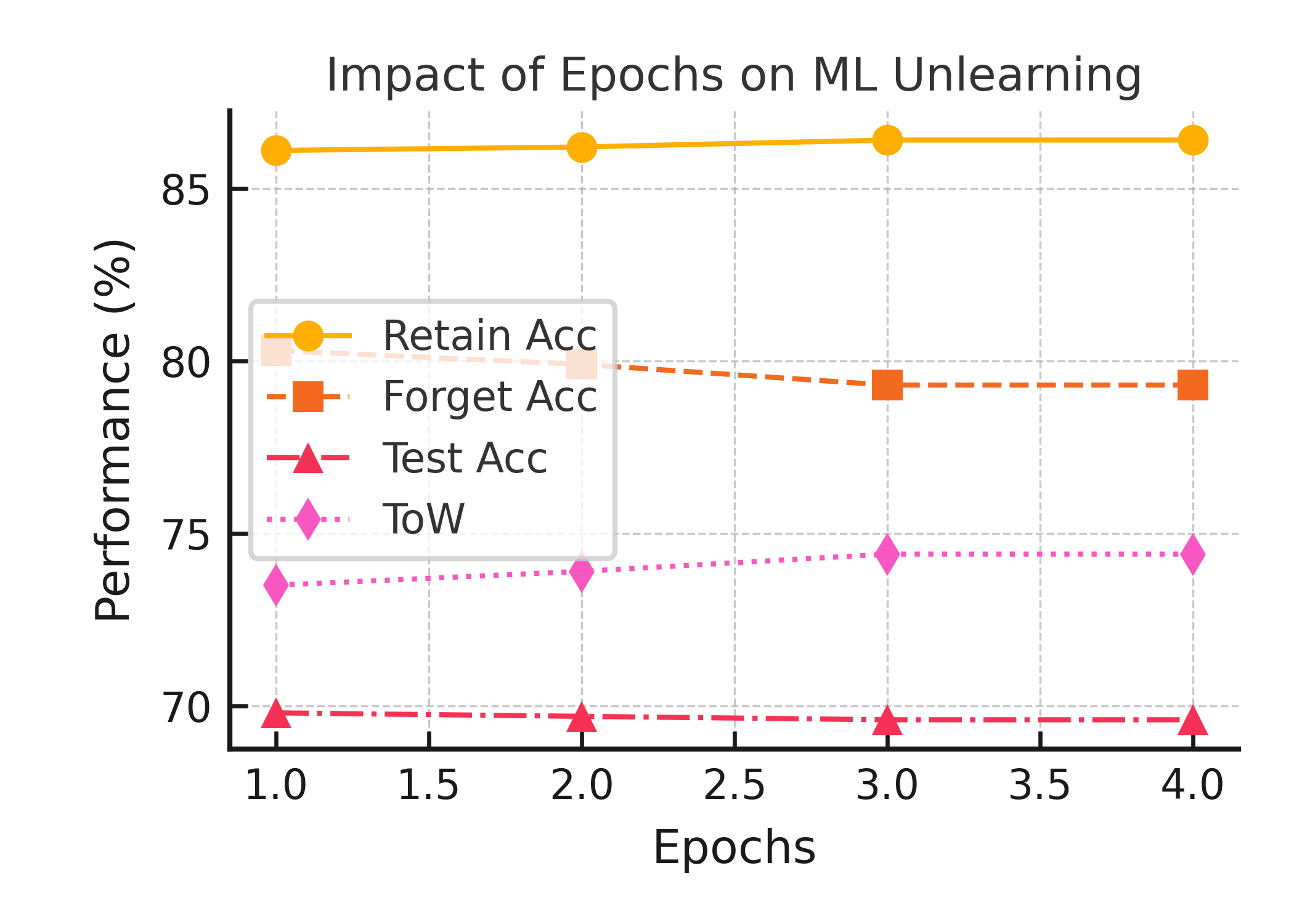}
    \caption{Impact of the number of readaptation epochs on
    downstream-task utility and membership-inference metrics.}
    \label{fig:impact-of-epochs-on-performance}
\end{figure}

\textbf{Sensitivity to Readaptation Epochs.}
We further analyze the effect of the number of readaptation epochs
$E$ on the MUFAC/10 setting. As shown in
Figure~\ref{fig:impact-of-epochs-on-performance} and
Table~\ref{tab:epochs_mia}, varying $E$ affects both downstream-task
utility and membership distinguishability.

Increasing the number of readaptation epochs initially changes the
utility--membership trade-off, whereas excessive readaptation can
degrade downstream-task utility in the evaluated setting
(e.g., $E>3$). Membership behavior is interpreted according to the
distance of the MIA AUC from $0.5$, rather than by assuming that a
monotonically smaller AUC necessarily indicates stronger privacy.

For the main experiments, we use $E=1$ to maintain low request-time
computational cost while providing a favorable empirical balance among
the reported utility and membership-inference measurements.

\begin{table}[ht]
    \centering
    \caption{Impact of the number of readaptation epochs on
    membership-inference AUC. Values closer to 0.5 indicate lower
    membership distinguishability under the evaluated attack.}
    \label{tab:epochs_mia}
    \begin{tabular}{cc}
        \toprule
        \textbf{Readaptation Epochs ($E$)} &
        \textbf{MIA AUC (ideal$\rightarrow$0.5)} \\
        \midrule
        1 & 0.5637 \\
        2 & 0.5512 \\
        3 & 0.5421 \\
        4 & 0.5387 \\
        \bottomrule
    \end{tabular}
\end{table}
% sec_discussion.tex

\section{Discussion}
\label{sec:discussion}
\textcolor{black}{
The experimental results demonstrate that MCU can provide a favorable
empirical trade-off among downstream-task utility, membership
distinguishability, and request-time computational efficiency under the
evaluated identity-level forgetting settings. Nevertheless, the current
study has several important limitations. We discuss these limitations
explicitly below and identify corresponding directions for future work.
\subsection{Limitations}
\label{subsec:limitations}
\textbf{Data availability and the scope of ``data-absent'' unlearning.}
MCU should not be interpreted as a fully data-free or retain-data-free
unlearning method. The external prototype memory is constructed from
representations derived from the original training data, and the
request-time model-readaptation stage uses both the one-shot support set
$\mathcal{D}_{\mathrm{support}}$ and the retain set
$\mathcal{D}_{r}$. The complete original forget set
$\mathcal{D}_{f}$, however, is not directly revisited during
gradient-based readaptation. Therefore, the data-absent setting
considered in this work specifically refers to the absence of the
complete original forget data during request-time gradient-based
readaptation, rather than the absence of all training-derived
information.
This distinction also means that the experimental setting adopts the
identity partitioning and one-shot support construction of the
1-SHUI benchmark~\cite{de2025unlearning}, but does not constitute a
strict support-only implementation of the original benchmark.
Accordingly, comparisons with other methods should be interpreted in
light of their potentially different request-time data-access
assumptions.
\\
\textbf{Privacy of external prototypes.}
Although MCU avoids directly revisiting the complete original forget
set during model readaptation, its identity prototypes are derived from
representations of the original training samples. These prototypes
should therefore be regarded as privacy-sensitive derived
representations rather than as privacy-neutral metadata. The present
study does not evaluate whether an adversary with access to the external
memory could infer attributes, reconstruct information about the
underlying samples, or perform prototype-inversion attacks.
Consequently, prototype deletion provides an explicit mechanism for
removing an identity-associated memory entry, but should not by itself
be interpreted as a formal guarantee that all information associated
with the forgotten identity has been eliminated from the entire model.
\\
\textbf{Scope of the privacy evaluation.}
The current privacy evaluation focuses on the Relativity-based
membership inference attack~\cite{zarifzadeh2023low}. An MIA AUC close
to $0.5$ indicates low membership distinguishability under this
specific evaluated attack, but it does not establish statistical
indistinguishability or a general privacy guarantee. Moreover, AUC
values substantially above or below $0.5$ may both indicate membership
separability depending on the orientation of the attack score.
The present evaluation does not include adaptive MCU-aware membership
inference attacks, model inversion, feature reconstruction, prototype
inversion, or other attacks specifically designed with knowledge of
the external-memory mechanism. The privacy conclusions of this study
are therefore restricted to the evaluated attack model.
\\
\textbf{Relationship to biometric applications.}
The evaluated datasets contain facial images and identity annotations,
but the downstream prediction tasks considered in this work are
facial-attribute prediction for CelebA and CelebA-HQ and age-group
classification for MUFAC. Identity serves as the unit of forgetting,
whereas the downstream labels define the prediction task.
Consequently, the current experiments should not be interpreted as an
evaluation of an operational biometric face-verification system based
on enrollment templates, embedding similarities, or verification
thresholds. Additional evaluation is required before extending the
observed results to such biometric settings.
\\
\textbf{Dependence on external-memory preparation and model
architecture.}
MCU augments the host model with an external MHN prototype memory,
a Prototype Adaptation Network (PAN), and a Dependency Gating Unit (DGU).
These components must be prepared before request-time unlearning.
Therefore, MCU is not a purely parameter-space post-hoc method that can
be applied to an arbitrary deployed model without prior preparation.
The reported request-time SpeedUp also excludes the one-time prototype
construction stage and should consequently be interpreted as
request-time efficiency after prototype preparation rather than as
end-to-end lifecycle acceleration.
Although the underlying idea of associating samples with external
representative patterns may potentially be applicable beyond the
architectures and datasets evaluated in this study, such generalization
has not been empirically established. The present results are limited
to the evaluated ResNet-18-based image models.
\\
\textbf{Scale of forgetting and component-level analysis.}
The current experiments consider forgetting requests involving up to
50 identities. The behavior of MCU when substantially larger fractions
of the stored identities are removed has not yet been evaluated.
Removing a large number of prototypes changes the remaining external
memory and may affect memory retrieval and downstream utility; this
possibility requires explicit investigation rather than being inferred
from the current experiments.
In addition, the present study does not provide a complete
component-wise ablation of the MHN, PAN, DGU, prototype-removal stage,
and individual readaptation losses. Although the complete MCU system
shows a favorable empirical trade-off in the reported experiments, the
individual contribution of each component cannot be isolated from the
current results.
\\
\textbf{Sensitivity to limited target-specific support information.}
The experimental protocol uses one held-out support sample for each
forget identity. MCU therefore operates with highly limited
target-specific information. However, the current study does not
systematically examine different numbers or qualities of support
samples. The robustness of prototype identification and readaptation to
noisy, atypical, or incorrectly selected support samples remains an
open question.
\subsection{Future Work}
\label{subsec:future_work}
Several extensions follow directly from the above limitations.
\textbf{Component-wise and data-access analysis.}
Future work will conduct controlled ablation studies to quantify the
individual contributions of the MHN, PAN, DGU, explicit prototype
removal, dependency-inducing loss, and retain-utility loss. It will
also examine the sensitivity of MCU to the amount of retain data and
to the number of target-specific support samples available at
request time. Such experiments would provide a clearer characterization
of the trade-off between data availability, forgetting behavior, and
retained utility.
\\
\textbf{Adaptive privacy evaluation and prototype protection.}
A broader security evaluation should include adaptive MCU-aware
membership inference attacks, prototype inversion, model inversion,
feature reconstruction, and attacks targeting the external-memory
interface. An important complementary direction is to investigate
privacy-preserving protection for the prototype store itself, such as
access-control mechanisms, protected execution environments, or
privacy-preserving prototype representations.
\\
\textbf{Large-scale and sequential forgetting.}
Future experiments will investigate substantially larger forgetting
requests as well as repeated or sequential forgetting operations.
These settings are particularly important for understanding how
progressive prototype removal changes MHN retrieval behavior and
whether repeated readaptation introduces cumulative degradation in
retain-set or test-set utility.
\\
\textbf{Support-sample robustness and pattern completion.}
Because the current protocol provides only one target-specific support
sample per forget identity, future work will evaluate sensitivity to
support-sample quality and quantity. The associative retrieval
capability of the MHN may also be explored as a mechanism for improving
the robustness of target-prototype identification when the available
support sample is noisy or atypical. This possibility, however, remains
to be experimentally validated.
\\
\textbf{From hard removal to controlled forgetting.}
The current MCU implementation removes the selected prototype from the
external memory. Future work could investigate more gradual forms of
memory modification, such as attenuating selected prototype
contributions, modifying retrieval temperature parameters, or
introducing mechanisms that suppress specific memory associations.
Such ``soft forgetting'' mechanisms may enable finer control over the
trade-off between forgetting strength and retained utility, although
their security properties would require separate evaluation.
\\
\textbf{Extension to operational biometric and non-image settings.}
Further evaluation is required before applying MCU to operational
biometric verification systems. Such studies should consider identity
embeddings, enrollment templates, similarity-based decision rules, and
verification-specific security metrics. Beyond facial-image tasks,
future work may also investigate whether the external-memory
unlearning principle transfers effectively to other modalities, such
as audio and text. These extensions should be treated as empirical
research questions rather than assumed consequences of the current
image-based results.
\\
\textbf{Toward auditable unlearning.}
Finally, future work may investigate mechanisms for auditing whether a
requested prototype has been removed and whether the subsequent model
state satisfies predefined empirical forgetting criteria. Secure
logging, trusted execution environments, or cryptographic techniques
may provide useful building blocks for such auditing mechanisms.
However, the current MCU framework does not provide a cryptographic or
formal proof of complete information removal.
}
\section{Conclusion}
\label{sec:conclusion}
\textcolor{black}{
This paper presented Memory-Coupled Unlearning (MCU), an
identity-level machine unlearning framework that augments a host model
with an external Modern Hopfield Network (MHN)-based prototype memory.
MCU formulates forgetting as an explicit memory-management process:
identity-associated prototypes are removed from the external memory,
followed by lightweight model readaptation through the Prototype Adaptation Network (PAN), Dependency Gating Unit (DGU), and the host
network. During request-time readaptation, MCU uses the one-shot support
set $\mathcal{D}_{\mathrm{support}}$ together with the retain set
$\mathcal{D}_{r}$, while the complete original forget set
$\mathcal{D}_{f}$ is not directly revisited. Therefore, MCU should not
be interpreted as a fully data-free or retain-data-free method; rather,
it reduces direct dependence on the complete target-specific forget
data during request-time gradient-based unlearning.
\\
Experiments on CelebA-HQ, CelebA, and MUFAC demonstrate that MCU
provides a favorable empirical trade-off among downstream-task utility,
membership distinguishability, and request-time computational
efficiency under the evaluated identity-level forgetting settings.
In particular, the results show that explicit prototype removal,
combined with retain-aware model readaptation, can preserve competitive
downstream utility while producing membership-inference behavior that
is close to random guessing in several evaluated settings. The reported
runtime results further show that MCU can substantially reduce
request-time unlearning cost relative to full retraining after the
external prototype memory has been prepared. We additionally use the
Utility-Privacy Balance Index (UPBI) as a complementary descriptive
metric for summarizing the observed utility--membership trade-off;
however, it is not intended to replace the individual utility and
attack-specific privacy measurements.
\\
More broadly, the results support the potential of treating machine
unlearning as an explicit memory-management problem rather than relying
solely on post-hoc parameter modification. By separating
identity-associated memory from the host model and enabling direct
prototype removal, MCU provides a controllable mechanism for
identity-level forgetting while retaining the ability to readapt the
model to preserve useful downstream behavior. Nevertheless, the current
results do not constitute a formal guarantee of complete information
removal. Further work is required to evaluate adaptive attacks,
prototype privacy, component-level contributions, larger and sequential
forgetting requests, and operational biometric verification settings.
These directions are important for assessing the broader applicability
and security properties of memory-coupled unlearning.
}
\section*{Generative AI Usage}

The authors disclose that LLMs were used for editorial purposes in this manuscript, specifically for grammar polishing and stylistic refinement of existing text drafted by the authors. All AI-generated outputs were manually inspected and verified by the authors to ensure accuracy and originality.

The core methodology, including the Memory-Coupled Unlearning (MCU) framework, its algorithms, experimental design, and all resulting analyses, are the original intellectual contributions of the authors and were not generated by an LLM.

\bibliographystyle{IEEEtran}
\bibliography{citation}

@article{xu2024machine,
  author = {Xu, Yi},
  title = {Machine Unlearning for Traditional Models and Large Language Models: A Short Survey},
  journal = {arXiv preprint arXiv:2404.01206},
  year = {2024},
}

@article{de2025unlearning,
  author = {De Min, Thomas and Mancini, Massimiliano and Lathuili{\`e}re, St{\'e}phane and Roy, Subhankar and Ricci, Elisa},
  title = {Unlearning personal data from a single image},
  journal = {Transactions on Machine Learning Research},
  year = {2025},
}

@article{goldfield2020right,
  author = {Goldfield, Charlene C},
  title = {'{T}he Right to Be Forgotten' and Its Unintended Consequences to Intelligence Gathering},
  journal = {Florida Journal of International Law},
  year = {2020},
  volume = {32},
  pages = {183},
}

@article{shi2024muse,
  author = {Shi, Weijia and Lee, Jaechan and Huang, Yangsibo and Malladi, Sadhika and Zhao, Jieyu and Holtzman, Ari and Liu, Daogao and Zettlemoyer, Luke and Smith, Noah A and Zhang, Chiyuan},
  title = {{MUSE}: Machine unlearning six-way evaluation for language models},
  journal = {arXiv preprint arXiv:2407.06460},
  year = {2024},
}

@article{yao2024machine,
  author = {Yao, Jin and Chien, Eli and Du, Minxin and Niu, Xinyao and Wang, Tianhao and Cheng, Zezhou and Yue, Xiang},
  title = {Machine unlearning of pre-trained large language models},
  journal = {arXiv preprint arXiv:2402.15159},
  year = {2024},
}

@article{teixeira2023can,
  author = {Teixeira da Silva, Jaime A and Nazarovets, Serhii},
  title = {Can the principle of the ‘right to be forgotten’ be applied to academic publishing? Probe from the perspective of personal rights, archival science, open science and post-publication peer review},
  journal = {Learned Publishing},
  year = {2023},
  volume = {36},
  number = {4},
  pages = {651--666},
}

@article{novikovas2022employee,
  author = {Novikovas, Andrejus and Grigonien{\.e}, Rasa},
  title = {Employee data retention periods in implementing the right to be forgotten: the situation in Lithuania},
  journal = {Entrepreneurship and Sustainability Issues},
  year = {2022},
  volume = {10},
  number = {2},
  pages = {623},
}

@article{correia2021right,
  author = {Correia, M{\'o}nica and Rego, Guilhermina and Nunes, Rui},
  title = {The right to be forgotten and COVID-19: privacy versus public interest.},
  journal = {Acta Bioethica},
  year = {2021},
  volume = {27},
  number = {1},
  pages = {59--67},
}

@article{kelly2021achieve,
  title={How to achieve compliance with gdpr article 17 in a hybrid cloud environment},
  author={Kelly, Miriam and Furey, Eoghan and Curran, Kevin},
  journal={Sci},
  volume={3},
  number={1},
  pages={3},
  year={2021},
  publisher={MDPI}
}

@inproceedings{kelly2019gdpr,
  title={GDPR Article 17: Eradicating personal identifiable information \& achieving compliance in a hybrid cloud},
  author={Kelly, Miriam and Furey, Eoghan and Blue, Juanita},
  booktitle = {Proceedings of the Irish Signals and Systems Conference},
  pages={1--6},
  year={2019},
  organization={IEEE}
}

@mastersthesis{zhen2022existence,
  author = {Zhen, Yue},
  title = {Existence and enforcement of a right to be forgotten in {China}},
  school = {Chulalongkorn University},
  year = {2022},
}

@inproceedings{thudi2022necessity,
  author = {Thudi, Anvith and Jia, Hengrui and Shumailov, Ilia and Papernot, Nicolas},
  title = {On the necessity of auditable algorithmic definitions for machine unlearning},
  booktitle = {Proceedings of the USENIX Security Symposium},
  year = {2022},
  pages = {4007--4022},
}

@inproceedings{eisenhofer2025verifiable,
  author = {Eisenhofer, Thorsten and Riepel, Doreen and Chandrasekaran, Varun and Ghosh, Esha and Ohrimenko, Olga and Papernot, Nicolas},
  title = {Verifiable and provably secure machine unlearning},
  booktitle = {Proceedings of the IEEE Conference on Secure and Trustworthy Machine Learning},
  year = {2025},
  pages = {479--496},
}

@inproceedings{bourtoule2021machine,
  author = {Bourtoule, Lucas and Chandrasekaran, Varun and Choquette-Choo, Christopher A and Jia, Hengrui and Travers, Adelin and Zhang, Baiwu and Lie, David and Papernot, Nicolas},
  title = {Machine unlearning},
  booktitle = {Proceedings of the IEEE Symposium on Security and Privacy},
  year = {2021},
  pages = {141--159},
}

@article{kurmanji2023towards,
  author = {Kurmanji, Meghdad and Triantafillou, Peter and Hayes, Jamie and Triantafillou, Eleni},
  title = {Towards unbounded machine unlearning},
  journal = {Advances in Neural Information Processing Systems},
  year = {2023},
  volume = {36},
  pages = {1957--1987},
}

@inproceedings{shi2025deepclean,
  author = {Shi, Jialei and Gourgoulias, Kostis and Buford, John F and Moran, Sean J and Ghalyan, Najah},
  title = {Deepclean: machine unlearning on the cheap by resetting privacy sensitive weights using the fisher diagonal},
  booktitle = {Proceedings of the European Conference on Computer Vision},
  year = {2025},
  pages = {1--16},
}

@article{foster2024loss,
  author = {Foster, Jack and Schoepf, Stefan and Brintrup, Alexandra},
  title = {Loss-free machine unlearning},
  journal = {arXiv preprint arXiv:2402.19308},
  year = {2024},
}

@article{yamashita2023one,
  author = {Yamashita, Tomoya and Yamada, Masanori and Shibata, Takashi},
  title = {One-Shot Machine Unlearning with Mnemonic Code},
  journal = {arXiv preprint arXiv:2306.05670},
  year = {2023},
}

@article{iacovelli2022gwfast,
  author = {Iacovelli, Francesco and Mancarella, Michele and Foffa, Stefano and Maggiore, Michele},
  title = {{GWFAST}: a Fisher information matrix python code for third-generation gravitational-wave detectors},
  journal = {The Astrophysical Journal Supplement Series},
  year = {2022},
  volume = {263},
  number = {1},
  pages = {2},
}

@article{neilsen2018optimal,
  author = {Neilsen, Tracianne B and Transtrum, Mark K and Van Komen, David F and Knobles, David P},
  title = {Optimal experimental design for machine learning using the Fisher information matrix},
  journal = {The Journal of the Acoustical Society of America},
  year = {2018},
  volume = {144},
  number = {3\_Supplement},
  pages = {1730--1730},
  publisher = {Acoustical Society of America},
}

@inproceedings{foster2024fast,
  author = {Foster, Jack and Schoepf, Stefan and Brintrup, Alexandra},
  title = {Fast machine unlearning without retraining through selective synaptic dampening},
  booktitle = {Proceedings of the AAAI Conference on Artificial Intelligence},
  year = {2024},
  volume = {38},
  pages = {12043--12051},
}

@inproceedings{zarifzadeh2023low,
  author = {Zarifzadeh, Sajjad and Liu, Philippe and Shokri, Reza},
  title = {Low-cost high-power membership inference by boosting relativity},
  booktitle = {Proceedings of the International Conference on Machine Learning},
  year = {2024},
  pages = {57749--57768},
}

@inproceedings{hu2024lwywtu,
  author = {Hu, Hongsheng and Wang, Shuo and Dong, Tian and Xue, Minhui},
  title = {Learn What You Want to Unlearn: Unlearning Inversion Attacks against Machine Unlearning},
  booktitle = {Proceedings of the IEEE Symposium on Security and Privacy},
  year = {2024},
  pages = {3257--3275},
  doi = {10.1109/SP54263.2024.00248},
}

@article{fan2023salun,
  author = {Fan, Chongyu and Liu, Jiancheng and Zhang, Yihua and Wong, Eric and Wei, Dennis and Liu, Sijia},
  title = {{SalUn}: Empowering machine unlearning via gradient-based weight saliency in both image classification and generation},
  journal = {arXiv preprint arXiv:2310.12508},
  year = {2023},
}

@inproceedings{chen2023boundary,
  author = {Chen, Min and Gao, Weizhuo and Liu, Gaoyang and Peng, Kai and Wang, Chen},
  title = {Boundary unlearning: Rapid forgetting of deep networks via shifting the decision boundary},
  booktitle = {Proceedings of the IEEE/CVF Conference on Computer Vision and Pattern Recognition},
  year = {2023},
  pages = {7766--7775},
}

@inproceedings{lee2023undo,
  author = {Lee, Sangyong and Woo, Simon S},
  title = {Undo: Effective and accurate unlearning method for deep neural networks},
  booktitle = {Proceedings of the ACM International Conference on Information and Knowledge Management},
  year = {2023},
  pages = {4043--4047},
}

@inproceedings{shrivastava2024refining,
  author = {Shrivastava, Alekh and Ahirwal, Mitul Kumar},
  title = {Refining of Learning in Human Decision Making Models: A Simple Step towards Machine Unlearning},
  booktitle = {Proceedings of the International Conference on Integrated Circuits, Communication, and Computing Systems},
  year = {2024},
  volume = {1},
  pages = {1--6},
}

@article{chen2024machine,
  author = {Chen, Huiqiang and Zhu, Tianqing and Yu, Xin and Zhou, Wanlei},
  title = {Machine unlearning via null space calibration},
  journal = {arXiv preprint arXiv:2404.13588},
  year = {2024},
}

@inproceedings{lin2023erm,
  author = {Lin, Shen and Zhang, Xiaoyu and Chen, Chenyang and Chen, Xiaofeng and Susilo, Willy},
  title = {{ERM-KTP}: Knowledge-level machine unlearning via knowledge transfer},
  booktitle = {Proceedings of the IEEE/CVF Conference on Computer Vision and Pattern Recognition},
  year = {2023},
  pages = {20147--20155},
}

@article{ramsauer2020hopfield,
  author = {Ramsauer, Hubert and Sch{\"a}fl, Bernhard and Lehner, Johannes and Seidl, Philipp and Widrich, Michael and Adler, Thomas and Gruber, Lukas and Holzleitner, Markus and Pavlovi{\'c}, Milena and Sandve, Geir Kjetil and others},
  title = {Hopfield networks is all you need},
  journal = {arXiv preprint arXiv:2008.02217},
  year = {2020},
}

@inproceedings{liu2015deep,
  author = {Liu, Ziwei and Luo, Ping and Wang, Xiaogang and Tang, Xiaoou},
  title = {Deep learning face attributes in the wild},
  booktitle = {Proceedings of the IEEE International Conference on Computer Vision},
  year = {2015},
  pages = {3730--3738},
}

@article{karras2017progressive,
  author = {Karras, Tero and Aila, Timo and Laine, Samuli and Lehtinen, Jaakko},
  title = {Progressive growing of gans for improved quality, stability, and variation},
  journal = {arXiv preprint arXiv:1710.10196},
  year = {2017},
}

@article{choi2023towards,
  author = {Choi, Dasol and Na, Dongbin},
  title = {Towards machine unlearning benchmarks: Forgetting the personal identities in facial recognition systems},
  journal = {arXiv preprint arXiv:2311.02240},
  year = {2023},
}

@article{zhao2024makes,
  title={What makes unlearning hard and what to do about it},
  author={Zhao, Kairan and Kurmanji, Meghdad and B{\u{a}}rbulescu, George-Octavian and Triantafillou, Eleni and Triantafillou, Peter},
  journal={Advances in Neural Information Processing Systems},
  volume={37},
  pages={12293--12333},
  year={2024}
}

@inproceedings{hoang2024learn,
  author = {Hoang, Tuan and Rana, Santu and Gupta, Sunil and Venkatesh, Svetha},
  title = {Learn to unlearn for deep neural networks: Minimizing unlearning interference with gradient projection},
  booktitle = {Proceedings of the IEEE/CVF Winter Conference on Applications of Computer Vision},
  year = {2024},
  pages = {4819--4828},
}

@article{foster2024zero,
  author = {Foster, Jack and Fogarty, Kyle and Schoepf, Stefan and {\"O}ztireli, Cengiz and Brintrup, Alexandra},
  title = {Zero-shot machine unlearning at scale via lipschitz regularization},
  journal = {arXiv e-prints},
  year = {2024},
  pages = {arXiv--2402},
}

@inproceedings{chundawat2023can,
  author = {Chundawat, Vikram S and Tarun, Ayush K and Mandal, Murari and Kankanhalli, Mohan},
  title = {Can bad teaching induce forgetting? unlearning in deep networks using an incompetent teacher},
  booktitle = {Proceedings of the AAAI Conference on Artificial Intelligence},
  year = {2023},
  volume = {37},
  pages = {7210--7217},
}

@inproceedings{shokri2017membership,
  author = {Shokri, Reza and Stronati, Marco and Song, Congzheng and Shmatikov, Vitaly},
  title = {Membership Inference Attacks Against Machine Learning Models},
  booktitle = {Proceedings of the IEEE Symposium on Security and Privacy},
  year = {2017},
  pages = {3--18},
  doi = {10.1109/SP.2017.41},
}

@book{GDPR,
  author = {Voigt, Paul and von dem Bussche, Axel},
  title = {The {EU} General Data Protection Regulation ({GDPR}): A Practical Guide},
  year = {2017},
  publisher = {Springer},
  doi = {10.1007/978-3-319-57959-7},
}

@article{RTBF,
  author = {Fosch-Villaronga, Eduard and Kieseberg, Peter and Li, Tiffany},
  title = {Humans forget, machines remember: Artificial intelligence and the right to be forgotten},
  journal = {Computer Law \& Security Review},
  year = {2018},
  volume = {34},
  number = {2},
  pages = {304--313},
}

@inproceedings{hayes2025inexact,
  author = {Hayes, Jamie and Khalifa, Amr and Shumailov, Ilia and Papernot, Nicolas and Triantafillou, Eleni},
  title = {Inexact Unlearning Needs More Careful Evaluations to Avoid a False Sense of Privacy},
  booktitle = {Proceedings of the IEEE Conference on Secure and Trustworthy Machine Learning},
  year = {2025},
  doi = {10.1109/SaTML64287.2025.00034}
}

@inproceedings{naderloui2025rectifying,
  author = {Naderloui, Nima and Yan, Shenao and Wang, Binghui and Fu, Jie and Wang, Wendy Hui and Liu, Weiran and Hong, Yuan},
  title = {Rectifying Privacy and Efficacy Measurements in Machine Unlearning: A New Inference Attack Perspective},
  booktitle = {Proceedings of the USENIX Security Symposium},
  year = {2025}
}
\vfill
\end{document}